# Self-induced spin glass state in elemental and crystalline neodymium

Umut Kamber[1], Anders Bergman[2], Andreas Eich[1], Diana Iuşan[2], Manuel Steinbrecher[1], Nadine Hauptmann[1], Lars Nordström[2], Mikhail I. Katsnelson[1], Daniel Wegner[1,*], Olle Eriksson[2,3], Alexander A. Khajetoorians[1,*]

1. Institute for Molecules and Materials, Radboud University, Nijmegen, The Netherlands

2. Department of Physics and Astronomy, Uppsala University, Uppsala, Sweden

3. School of Science and Technology, Örebro University, SE-701 82 Örebro, Sweden

*corresponding authors: a.khajetoorians@science.ru.nl, d.wegner@science.ru.nl

**Spin glasses are a highly complex magnetic state of matter, intricately linked to spin frustration and structural disorder. They exhibit no long-range order and exude aging phenomena, distinguishing them from quantum spin liquids. We report a new type of spin glass state, the spin-Q glass, observable in bulk-like crystalline metallic neodymium thick films. Using spin-polarized scanning tunneling microscopy combined with ab initio calculations and atomistic spin-dynamics simulations, we visualize the variations in atomic-scale non-collinear order and its response to magnetic field and temperature. We quantify the aging phenomena relating the glassiness to crystalline symmetry and the energy landscape. This result not only resolves the long-standing debate of the magnetism of neodymium, but also suggests that glassiness may arise in other magnetic solids lacking extrinsic disorder.**

Spin glasses are one of the more intriguing, but least understood magnetic states of matter (*1-5*). Ferromagnets or antiferromagnets form a long-range ordered state when cooled, but spin glasses form a state characterized by seemingly random and uncorrelated magnetic patterns. The magnetization pattern in spin glasses can be compared to the amorphous structure of glasses like silicon dioxide that exhibit local structural correlations but lack a long-range ordered state. Interest in spin glasses spans many fields, ranging from iron-based superconductors (*6*) to theoretical machine learning (*4, 5, 7*), and it has been suggested to be relevant in quantum topological excitations.

Spin glasses are characterized by a glass transition temperature and by aging, that is, the magnetic state depends on its history, driven by a distribution of distinctive spin-relaxation processes with time scales spanning many orders of magnitude (*4, 5*). Aging also distinguishes spin glasses from so-called quantum spin liquids (*8*) or spin ices (*9*), which remain disordered down to zero temperature because of quantum fluctuations and lack of memory. The paradigm of all of these types of complex magnets involves magnetic frustration, derived from geometry or competing interactions. However, unlike spin liquids, disorder is traditionally also considered necessary to drive non-ergodic behavior in spin glasses.

There is still no clear understanding when spin glass behavior can arise in magnetic materials. The most commonly debated models (*1, 3-5*) describe randomly distributed spins with long-range magnetic interactions of alternating ferromagnetic and antiferromagnetic coupling, like that seen in prototypical materials such as dilute magnetic alloys (*1, 10*). In the thermodynamic limit, spin glasses have a hierarchical energy landscape with infinitely many local energy minima separated by energy barriers of multiple heights so that there is a broad distribution of transition times between different minima (*3-5, 11*) that results in an absence of local and long-range order. Most models of spin glasses invoke structural disorder, i.e. amorphousness, as a key requirement along with competing magnetic interactions (*12*). However, self-induced glassiness, a concept introduced initially for the stripe glass behavior in high-temperature superconductors (*13, 14*), was later developed for magnets (*15, 16*). Within this framework, competing interactions alone can lead to the glassy state, even in the absence of external disorder, as well as can give rise to intermediate regimes which exhibit multi-well potentials (*17*).

We show that single-crystalline elemental neodymium exhibits a new type of spin glass behavior. Spin-polarized scanning tunneling microscopy on the (0001) surface of thick Nd films revealed that the magnetic state exhibited strong local non-collinear magnetic order while lacking a long-range ordered state. This local order is defined by a spectral distribution of degenerate magnetic wave vectors, or Q states, which varied spatially and with time. We probed the response of this so-called spin-Q glass to applied magnetic fields and variable temperature, to quantify its aging behavior and energy landscape. Harnessing ab initio methods and simulations, we quantified the competing long-range magnetic interactions and the favorable Q states, illustrating that this unconventional glassy behavior results from valley-like pockets of degenerate Q states as proposed for self-induced spin glasses (*15, 16*). Moreover, we performed calculations of the autocorrelation function for pristine neodymium which also found multiple relaxation times in its spin dynamics. Our findings not only suggest that glassy behavior and aging can be found in systems with crystalline order, but also unravels an unresolved debate about the magnetic ground state of elemental Nd that has challenged scientists for several decades.

**The magnetic ground state of Nd(0001)**

Despite more than 50 years of investigations, there is still no consensus on the magnetic ground state of Nd and the origin of the complexity reported in various experimental observations (*18-26*). Below the Néel temperature ($T_N$), neutron diffraction observed the onset of static multi-Q states with decreasing temperature. Yet, it is not well understood how the multiplicity of these Q states depends on the exchange landscape of the material and which real-space magnetic interactions causes them. Moreover, other measurements indicated additional phase transitions below $T_N$, suggesting that the original conclusions from neutron diffraction of a modulated antiferromagnetic structure were oversimplified (*23*). In addition, there are experimental observations above $T_N$ that give evidence for short-range order as well as a strongly frustrated exchange landscape (*26, 27*). These open questions illustrate a need for a characterization of the exchange interactions in Nd, as well as a real-space characterization of the local magnetic order.

**Atomic-scale visualization of the spin-Q glass state**

To characterize the atomic-scale magnetization of the surface of Nd(0001), we utilize spin-polarized scanning tunneling microscopy and spectroscopy (SP-STM/SP-STS) (*28*). We epitaxially grow thick films (up to 100 monolayers (ML)) of Nd(0001)/W(110). We note that lanthanide films prepared in this way on various bcc(110) substrates exhibit high crystallinity and superior surface cleanliness over sputter-annealed bulk single crystals (*29-37*). It has been shown for various lanthanide elements, including Nd, that films grow with bulk lattice parameters above a thickness of 10 ML (*31, 35, 38-40*), exhibit a bulk-like electronic structure above 30 ML (*32, 33, 41*), and start to exhibit bulk-like magnetic behavior in the range of 30-50 ML (*42-45*). Superlattice studies showed that 33-39 ML-thick Nd films already exhibit the temperature-dependent neutron scattering features as known from bulk single crystals (*38, 46, 47*).

The sample morphology can be tuned to either layer-by-layer grown closed films or islands (Stranski-Krastanov (SK) growth), depending on the annealing temperature (*39, 42, 48-50*). We grew two types of samples: SK grown islands of >50 ML thickness (Fig. 1A,B) and closed ~100 ML films (Fig. S1B) (*50*). As discussed above, SK islands should readily represent bulk-like structural, electronic and magnetic properties. We verified this on thicker (~100 ML) closed films, which showed identical magnetization patterns as we subsequently discuss (see supplementary text, section S2) (*50*). However, we observe that closed thick films show inferior surface qualities, due to the presence of more impurities as well as screw dislocations that can lead to pinning of magnetic structures (*51*). As the latter was not present for our island-grown samples (Fig. S1A), we focus on data taken from islands with thickness between 58 and 92 ML and lateral sizes between 58000 and 200000 nm$^2$, and may account for differences in previous measurements. Further details of the sample preparation and morphology as well as a discussion of thin films vs. bulk samples can be found in the supplementary text, section S1 (*50*).

Using scanning tunneling spectroscopy (STS) at low temperature (0.03−7 K), we probed the Nd(0001) surface state, whose presence and sharpness has been shown to be a probe of the cleanliness of the film (*27, 31, 32, 52, 53*). The surface state is characterized by an exchange splitting into a majority peak visible below the Fermi energy ($E_F$, $V_S$ = 0) and a minority peak above $E_F$, with an additional narrow peak at $E_F$ (Fig. 1C) (*49*). We note that we saw no difference between spectra taken on the islands and on locally clean areas of the thick film. Furthermore, the magnetic exchange splitting reproduced previous low-temperature data taken on a 30 ML film (*27, 54*), which is further evidence that our samples are beyond the thin-film limit and fully reflect bulk electronic and magnetic properties.

In order to get magnetic contrast, we utilized the spin-polarized nature of the exchange-split surface state, and we imaged the surface in constant-current mode at two characteristic voltages representing dominant tunneling into the minority state ($V_S$ = 200 mV) and out of the majority state ($V_S$ = -150 mV), respectively (Fig. S2) (*50*). We used an out-of-plane sensitive antiferromagnetic Cr probe, which relates spin-dependent contrast variations directly to the $z$-projection, that is, the $c$-axis in the double hexagonally close packed (dhcp) structure, of the magnetization. We consider the subtracted image in order to remove stronger topographic modulations resulting from buried substrate steps or locally varying sample thickness. We refer to this subtracted image as the magnetization image (Fig. 1D,G and Fig. S2D). More details on this image processing procedure can be found in the supplementary text, section S2) (*50*).

The magnetization images of the surface revealed strong and clear short-range magnetic order with periodicities $\lambda$ varying from 0.9 to 4.5 nm and oriented along or near the high-symmetry axes. These atomic-scale variations are directly related to local non-collinear magnetic order, with varying periodicities depending on spatial location, defined by a superposition of local magnetic wave vectors $Q_i = 2\pi/\lambda_i$ (Fig. 1G,H; supplementary text, section S3) (*50*). Remarkably, although clear short-range order could be seen, defined by a local multi-Q state, there was no observable long-range ordered state found for any of the probed experimental conditions.

In order to better visualize the variations of local multi-Q order, we consider reciprocal-space images obtained through Fast Fourier Transform (FFT) of the real-space magnetization images, which we refer to as Q-space images. A signature of the lack of long-range order and competing short-range order can be directly visualized by the smeared and broadly distributed spectral weight in various regions in Q space (Fig. 1E). We note that Q vectors here were derived from real-space SP-STM maps, and not directly measured as in the case of neutron diffraction (see supplementary text, section S3 for further discussion) (*50*). We only produced and analyzed Q-space images from within a flat terrace on individual islands (typical width ≈ 150 nm). Thus, the measured broadening resulted from the spectral distribution of Q, and not from averaging over different islands with different orientations.

To highlight this, Q-space images were also produced from smaller spatial regions of the same image, which illustrated sharpened and characteristic spectral weight of Q vectors compared to the larger scale image (Figs. 1G, S4). Various line-cuts along the $\bar{\Gamma}$-$\bar{M}$ direction of multiple Q-space images taken from different spatial regions of Fig. 1D are illustrated in Fig. 1F. The resultant plots revealed a spatial variation in spectral weight in at least three distinct regions, or Q-pockets, with substantial spectral weight along the high-symmetry axes.

In the ensuing discussion, we focus particularly on the three Q-pockets with wave vectors at around $Q_A$ = 1.1-2.0 nm$^{-1}$, $Q_B$ = 2.7-3.5 nm$^{-1}$ and $Q_C$ = 4.6-5.3 nm$^{-1}$. From this information, we plot a schematic of the out-of-plane projected magnetization with respect to the atomic lattice (Fig. 1H) for the two regions shown in Fig. 1G with the defined Q vectors from these Q-pockets. The corresponding Q-space images are a direct visualization of the multi-Q nature acquired over the spatial area of the given images, and they provide a quantitative comparison to previous neutron diffraction studies (*18-21, 24-26*). The spectral weight around these pockets, as well as blurring of the intensity of the Q states in larger scale images (Fig. 1E) was an initial indication of the glassy nature of the magnetic state, and reminiscent of spin-based analogs of stripe or checkerboard order in strongly correlated compounds (*55*).

In order to illuminate the concept of a spin-Q glass, we qualitatively illustrate the energy landscape in Q-space images for a spin-Q glass in comparison to a ferromagnet (Fig. 2). A long-range ordered state can be related to a global minimum in Q space (*28, 56, 57*), where a single-domain ferromagnetic state is equivalent to a Q = 0 global minimum (Fig. 2A). In contrast, a spin-Q glass is distinguished by the existence of flat valleys defined by a distribution of many local minima, i.e. Q-pockets, at finite Q values (Fig. 2B). Broad Q-pockets led to a lack of a preferentially long-range ordered state. Instead, there were local regions defined by strong local order derived from a spectral weight of mixed Q vectors within the given pockets, and different regions exhibited random distributions of this spectral weight (see color image in Fig. 2B). The superposition of this spatially varying magnetization led to an overall broadening in the spectral weight. Note that multi-Q states for thin 3*d* transition metal films show well-ordered domains (*58*). Within the concept of self-induced glassiness for spins, such pockets may result from a strong competition of magnetic interactions, leading to highly degenerate states (*15, 16*).

**Theoretical analysis of the magnetic landscape of Nd(0001)**

In order to analyze the origin of the spin-Q glass state in Nd(0001) and its unexpected magnetic patterns (*59*), we used ab initio calculations to quantify and understand the exchange interactions and the energy landscape in Q space of bulk Nd, which adopts a dhcp structure that is critical for its magnetic exchange interactions (Fig. 3A). The RSPt code was used for this purpose (see methods and supplemental material, section S4) (*50*). The calculated exchange interactions for bulk Nd are shown for both the dhcp and a hypothetical hcp structure and note the minute energy difference between these crystal structures, resulting from the large similarities in atomic arrangement.

The distance dependence of the exchange interactions in the hcp structure illustrates a prototypical behavior with ferromagnetic nearest neighbor interactions and an oscillating Ruderman-Kittel-Kasuya-Yosida (RKKY)–like interaction at large distances, as seen with other lanthanides such as Gd (*60, 61*). In contrast, in the dhcp, at shorter range, the interactions are much weaker and primarily antiferromagnetic, but at larger distances, an RKKY interaction sets in. These strong competing magnetic interactions in Nd create conditions for frustrated magnetism and spin-glass formation, as described in the picture of self-induced glassiness (*15-17*). Calculated values of the magnetic moment of the surface atoms, and for deeper layers, were similar to the bulk value. The bulk moment was restored within three layers beneath the surface (supplementary text, section S4) (*50*).

In order to clarify the impact of the calculated exchange interactions in the dhcp structure of Nd, we evaluated the magnetic energy landscape by means of single-Q spin spirals, i.e. magnetic structures that can be parametrized by a single wave vector. For this purpose, we calculated the energy of helical spin spirals. Using the calculated magnetic exchange interactions from Fig. 3A, we parametrized an effective Heisenberg spin Hamiltonian from which the energy $E(\mathbf{Q})$ of the single-Q spirals was then calculated (Fig. 3B). The energy-landscape exploration was performed by fixing the z-component $Q_z$ and then sweeping over $Q_x$ and $Q_y$ in the first Brillouin zone (BZ) (supplementary text, section S6) (*50*).

In Fig. 3B, we present the single-Q energy landscape for all possible $Q_x$ and $Q_y$ combinations in the cell spanned by the dhcp reciprocal lattice vectors for the case when $Q_z = 2\pi/c$, which corresponded to the configuration with the lowest single-Q energy. This choice of $Q_z$ corresponded to a 90° rotation of the moments in adjacent atomic layers within the dhcp unit cell. The Q-dependent energy is color-coded such that red (blue) regions correspond to spin spirals with low (high) energy. This visualization illustrates the complex energy landscape of Nd, as exemplified by the broad and flat dark-red ring-like structure, showing similarity to the deduced landscape in (*26*).

Instead of a distinct, six-fold degenerate set of strong energy minima that could be expected for a spin-spiral magnet on a hexagonal lattice, the red ring structure showed that the energy barriers between global and local minima in this region of the BZ were very small. In addition, high-energy local minima, or pockets (in red) were distributed with hexagonal symmetry, just inside the BZ, and another low-energy valley-like structure formed along the BZ boundary. Magnetic anisotropy was not considered here, and we expect that it would bias the ring structure toward the high-symmetry directions, creating pockets akin to those seen in the experiments. Thus, Fig. 3B illustrates that the energy landscape of Nd has several broad Q-pockets, which supports the formation of the experimentally observed spin-Q glass structure. The presence of strongly competing interactions leading to a glass-like energy landscape in Q space is a key manifestation of the concept of self-induced glassiness (*13-15*).

**Spin-Q glass dynamics: autocorrelation and static correlation function**

In addition to the single-Q energy-landscape explorations, we also used Monte Carlo and atomistic spin dynamics (ASD) simulations to find the ground-state magnetic structure. Although not all conventional spin glasses exhibit identical relaxation dynamics, a common trait is aging, i.e. that the relaxation process slows down over time and never settles on a single equilibrium state. Aging is observed in the Edwards-Anderson model (*1*). In order to characterize the aging dynamics of dhcp Nd we studied the two-time autocorrelation function defined as $C(t_w, t) = \langle \bm{m}_i(t + t_w) \cdot \bm{m}_i(t_w) \rangle$, where the brackets denote averaging over all sites of the system. If a system relaxes to a fixed ground state following non-glassy dynamics, the autocorrelation function should increase with increasing waiting times $t_w$. Autocorrelation analysis based on ASD has earlier been used to capture the multiple relaxation scales of a conventional spin glass system (*10*). This approach, which is typically analyzed with a mean-field approach, is well suited to characterize the multiple time scales indicative of aging (supplementary text, section S8) (*50*).

In Fig. 3C, we show the autocorrelation at logarithmically spaced waiting times $t_w$ for dhcp Nd as it relaxed from a fully disordered state at $T$ = 1 K. The autocorrelation function decayed exponentially towards zero, a typical feature of aging and spin-glass behavior (Fig. 3C) (*50*), and was similar to that of a traditional spin-glass system, like Cu-Mn or that of the Edwards-Anderson model (see supplementary text, section S8) (*10, 50*). In other words, the relaxation process of Nd never arrived at a well-defined energy minimum. We note the autocorrelation analysis often assumes a mean-field picture of aging, whereas a more complete picture should include the Q-dependent contributions to the multiple time scales. These simulations were performed for bulk dhcp Nd meaning that the observed spin-Q glassy behavior can only be a result of the exchange interactions of the system since neither surface states nor defects were present in the simulations.

In addition to mapping out the aging dynamics of the spin-Q glass state, from these simulations, we obtained real-space spin structures (supplementary text, section S7) (*50*) as well as the static correlation function $S(\mathbf{Q})$, which can be compared directly with the experimental Q-space images. In Fig. 3D, we present the simulated peaks of $S(\mathbf{Q})$ in comparison with (i) the observed range of spectral weight from the Q-pockets $Q_A$, $Q_B$ and $Q_C$ in the SP-STM experiments ($B_z$ = 0, for both pristine zero-field cooled samples and samples after being exposed to magnetic field), and (ii) reported neutron diffraction data (*21, 25*). In comparison, the SP-STM and neutron diffraction data agreed well with the simulated results in regions of lower Q values. This agreement indicated that the surface-derived measurements were significantly coupled to the bulk magnetic properties. The simulations also provide distinct local maxima of the correlation function, which so far have not been detected experimentally.

**Impact of impurities on Q-state distribution**

Nd films illustrate characteristics of bulk Nd with the dhcp structure, including the expected surface state of Nd(0001) and the expected strain-free lattice constant (*50*). Although we did not observe bulk screw dislocations and large-scale defects for the island samples, the surface did show the presence of surface impurities. The presence of the interface did not seem to influence the imaged magnetism, e.g. underlying substrate step edges (supplementary text, section S2) (*50*). Before discussing the role of the impurities on the magnetic order, we note that the main concentration of impurities in the source material is oxygen and carbon, each about 2000 parts per million by atom (ppma) (Table S1) (*50*). The impurity density of our surfaces is ca. 0.01 ML, which is below the detection limit of any surface-averaging spectroscopic technique and among the best reported for lanthanide metal surfaces (*53, 62*). Of these impurities, we observe the unperturbed intrinsic width of the surface state (*54*). As the quality of the surface state (i.e. its intensity and width) in spatially averaging photoemission has been demonstrated to illustrate the structural order as well as cleanliness of the surface (*31, 32, 52*), we conclude that the surface defects in the present limit have a negligible impact on the overall electronic structure.

In Fig. 4, we show a comparison between two separately prepared samples and their respective FFTs, in which the defect density is sufficiently different. For samples with higher defect densities, imaged at $T$ = 1.3 K, we observed more well-defined Q vectors at higher overall Q values, with a narrower distribution in Q space compared to samples where the defect density is lower (e.g. Fig. 1). Although local order in this case had a well-defined periodicity, the orientation was not perfectly aligned leading to a small angular distribution of Q states. In stark contrast, as the defect density was reduced, the distribution of Q states strongly broadened. In other words, with cleaner samples, there was a stronger manifestation of multiple energy minima as more Q states stabilize. This observation is in contrast to typical magnetically ordered systems, where any collinear or non-collinear configuration narrows its distribution in reciprocal space with cleaner samples. In this context, extend line defects (screw and edge dislocations) in lanthanide films were shown to pin magnetic order, while local point defects such as atomic adsorbates did not (*51*). As shown in Fig. S1 and S3 (*50*), thick closed Nd films exhibit line defects as well as higher surface defect densities than the typical islands we show in Fig. 1. The unavoidable presence of dislocations and the higher amount of impurities in bulk single crystals may lead to defect-induced pinning of Q-vectors, exemplifying the importance of cleanliness in order to observe the glassy behavior. However, for the ~100 ML thick closed films we observed qualitatively the same magnetic multi-Q structure as the islands (Fig. S3) (*50*). Hence, despite their seemingly small volume the islands reflect the magnetic properties of bulk Nd. We conclude that below a critical defect density, a multi-well landscape emerged as a spin-Q glass phase, as illustrated in the defect-free calculations of both the energy landscape and the autocorrelation function, in Fig. 3C.

**Magnetic field evolution of the spin-Q glass state**

We experimentally characterized the response of the magnetization to external magnetic fields. Out-of-plane field dependence is illustrated in Fig. 5 for a few chosen fields up to $B_z$ = 7 T at $T$ = 1.3 K (see also supplementary text, section S9) (*50*). The application of magnetic field should favor states with smaller Q, eventually leading to a preferential Q state near the zone center (*56, 63*), when the Zeeman energy exceeds the local exchange energy. However, magnetic imaging of a given area at variable magnetic field revealed a different picture. At increasing magnetic fields, no distinct Q state became favorable, with the spectral weight strongly broadening along high-symmetry directions toward higher Q (Fig. 5G). The magnetic order was sensitive to fields on the order of $B_z$ = 0.5 T, illustrating the degeneracy driven by the Q-pockets, whereas at the highest applied fields ($B_z$ = 7 T) there was no distinct Q state, demonstrating the strong local exchange energy.

Moreover, the application of in-plane fields revealed similar behavior: favorable Q states collapse onto an axis related to the direction of the applied field, but with appreciable smearing of the spectral weight along one particular axis, reminiscent of the so-called archipelago phase seen in neutron scattering (*22-24*) (Fig. S13) (*50*). The field-dependent behavior seen here cannot be attributed to the picture of local domains, as there were neither clear domain boundaries that could be traced in magnetic field, nor a repeating or favorable Q-structure. We further note that magnetostriction effects should saturate beyond 1 T, at the measured temperature, as previously reported (*23*).

**Aging of the spin-Q glass state through magnetic field**

Although field-dependent imaging illustrated the degeneracy within the Q-pockets, the distinguishing property of a spin glass is the observation of aging. Aging, in a mean-field picture, can be described by the existence of multiple relaxation time scales, as exemplified in Fig. 3C, leading to a magnetization state that never fully relaxes (*10*). To monitor the evolution of the ground state, we repeatedly ($i = 1, ..., n$) apply the following procedure, starting from the pristine ($i = 0$) state : (i) sweep up the magnetic field to $B_{z,i}$ (max. 7 T) and stay at that value for a finite time $\tau_i$, (ii) reduce the magnetic field to zero, (iii) image the $i^{th}$ zero-field state of the same area as the pristine state. Comparing several field-dependent cycles, the zero-field magnetic state at $T$ = 1.3 K illustrates sequential redistributions of the spectral weight within and between the various Q-pockets (Fig. 6).

Consistent with aging, the magnetic state did not relax to a given ground state after perturbation, nor was there a clear tendency toward a different but particular ground state distribution. Instead, the overall trend was a redistribution between all pockets toward a broader overall distribution along the high-symmetry directions, with the angular distribution sharpening along the high-symmetry axes. Moreover, the system never reverted to the initial zero-field cooled state. These findings ruled out that field-dependent cycling can be understood as the evolution of a favorable domain. To rule out hysteretic effects, we also performed field sweeps at positive and negative fields, but there was no clear correlation between such subsequent conditions, nor any correlation with local defects on the surface. We note that similar aging effects, and magnetic field evolution, were seen at different temperatures (Fig. S15) (*50*), as we detail below, as well as with the application of an in-plane field and also near island edges (Fig. S14, see also supplementary text, section S10 and S11) (*50*).

**Multiple and Q-dependent relaxation times at elevated temperature**

The evolution of intermittent spectral weight distributions, which were frozen after subsequent field-sweep cycles, is a signature of slow aging dynamics and a hierarchy of states separated by small energies. In contrast, the application of temperature illustrated the presence of fast dynamics concomitant with slow relaxation dynamics. To illustrate this effect, we show the effect of temperature by imaging the same area at different temperature (Fig. 7). Compared to $T$ = 1.3 K (Fig. 7B), the spectral weight within the $Q_A$ pockets "melted" away at $T$ = 4.2 K (Fig. 7D). Also in the real-space image, the associated long-wavelength pattern was no longer visible (Fig. 7C) when compared to images of the identical area at $T$ = 1.3 K (Fig. 7A). Our time resolution was limited to roughly 1 ms so that we were imaging the time-averaged magnetization. Thus, the loss of spectral intensity in the $Q_A$ pockets was most likely caused by increased fluctuations of the magnetization states within the $Q_A$ pockets, which were thermally activated at this temperature. Thus, the fluctuations of the magnetization lead to an overall reduction of the measured time-averaged spin polarization. Nevertheless, we observed similar slow aging behavior after application of magnetic field at $T$ = 4.2 K for the other Q-pockets as seen at lower temperatures for exactly the same sample area (Fig. 7F). This observation illustrated that there were at least two different and Q-dependent time scales present at sufficient temperature.

Although it is unclear how to describe the Q-dependent dynamical behavior of the magnetization, the presence of multiple relaxation times is a clue of glassy behavior in a mean-field picture, analogous to aging effects seen in metallic alloys (*3, 10*). We note that there may be some ruggedness to the energy barriers, symbolic of a multi-well landscape, resulting from strong local order in this system. In structural glasses, local order leads to a more complex picture of the dynamics, referred to as dynamic heterogeneity (*64*). In this picture, there can be different spatial regions exhibiting dramatically different dynamics related to the local correlations. The existence of different and Q-dependent time scales suggests evidence for dynamic heterogeneity in Nd. Therefore, it is not entirely clear how many various time scales can be seen in this system, and at what temperatures, compared to conventional spin glasses which have more flat energy landscapes. Combined with the observations that lower defect densities led to a smearing of the Q states, these results exemplify the complex aging behavior related to the multi-well energy landscape illustrated by theory in Fig. 3. The application of higher temperature led to more well-defined Q states (e.g. for $T$ = 7 K); the same area imaged at $T$ = 40 mK, in comparison, shows similar glassy distributions to the previously shown data taken at $T$ = 1.3 K (supplementary text, section S11) (*50*). This result rules out transitions into various long-range ordered multi-Q states (*21, 25*) with decreasing temperature, but rather exemplifies the emergence of a spin-Q glass state.

**Discussion**

We have demonstrated that the magnetic ground state of elemental Nd is a spin-Q glass. We have observed aging, that is, glassiness, in an elemental magnetic solid with minimal amounts of chemical or extrinsic disorder, which is direct experimental confirmation of self-induced glassiness (*15*). Moreover, the coexistence of short-range order exhibited by multi-Q pockets along with aging behavior in a material without extrinsic disorder cannot be captured by traditional mean-field theoretical descriptions of spin glasses (*1, 2*). The self-induced glassy behavior observed here is a departure from the traditional magnetic alloys where disorder drives the glassy dynamics and likewise many intermediate time scales can be observed. The energy landscape of Nd may possess some amount of ruggedness, resulting from the strong local Q order (*17*), providing a novel material system to study dynamic heterogeneity in a spin glass material (*64*). With the advent of techniques that can probe picosecond dynamics with STM (*65*), it may be interesting to develop magnetic sensitive time-resolved methods that can resolve the picosecond dynamics in neodymium, which can be directly compared to ASD simulations. It remains to be understood what is particularly special about the interplay between crystal structure and electronic properties in Nd that leads to this exotic behavior, necessitating a deeper theoretical understanding of the role of electron correlation effects, the possible influence of the surface, as well as the interplay between spin and orbital degrees of freedom.

The conclusions drawn here resolve the long-standing debate about the magnetic state of Nd, in that the various reported multi-Q transitions as a function of temperature can be understood within a picture of multiple minima in Q space with different depths that can be thermally activated. Likewise, the establishment of spin-Q glass order raises the question of the dynamical behavior of spins through the variation in local relaxation times, and may provide a material platforms to explore exotic topological quasiparticles similar to fractons (*66-68*). The example here expands our views on magnetic states of matter, necessitating numerous further experimental and theoretical investigation in order to understand the emergence of aging behavior in magnetic systems.

**Methods**

The experimental studies were performed in two home-built ultrahigh vacuum (UHV) systems that allow for cleaning the W(110) single crystal substrates, molecular beam epitaxy and annealing of Nd (see supplementary text, section S1) as well as transfer into the cryogenic STM, all in situ. The first system operates at a base temperature of 1.2 K with magnetic fields up to 9 T perpendicular to the sample. The second system operates at a base temperature of 30 mK by with magnetic fields up to 9 T perpendicular and 4 T parallel to the sample (*69*). The SP-STM measurements were performed using an antiferromagnetic Cr tip with out-of-plane spin polarization (see supplementary text, section S2). Apart from a global plane subtraction, all STM topography images are unprocessed raw data. Magnetization images were produced by subtraction of the majority and the minority SP-STM images, and corresponding Q-space images were produced by computing the FFT using MATLAB. For the latter, we again did not apply any post-processing (see supplementary text, section S2).

The theoretical studies included ab initio calculations of bulk Nd as well as cubic- and hexagonal-terminated slabs with up to 13 layers. The electronic structure was calculated within density functional theory by means of the RSPt software (*70*). We used the local spin-density approximation for the exchange and correlation functional. In these calculations we made use of the standard model of the lanthanides and treated the 4*f* electrons as localized, unhybridized with a magnetic moment according to Russel-Saunders coupling. The dispersive states were expanded by means of 6*s*, 6*p* and 5*d* orbitals in a multiple basis fashion, as described in (*70*). In addition, we included pseudo-core 5*s* and 5*p* states to hybridize with the rest of the valence states. No approximation was made concerning the shape of the charge density or potential, in a so-called full-potential description. The simulations were done using 32768 and 1152 k-points of the full Brillouin-zone for bulk dhcp and slabs, respectively (see also supplementary text, section S4). The MC/ASD calculations were performed using the UppASD software (*71, 72*) (see supplementary text, section S5).

**Acknowledgements**

**Funding:** We would like to acknowledge the Swedish National Infrastructure for Computing (SNIC). We also acknowledge funding from NWO, and the VIDI project: "Manipulating the interplay between superconductivity and chiral magnetism at the single-atom level" with project number 680-47-534. This project has received funding from the European Research Council (ERC) under the European Union's Horizon 2020 research and innovation programme (SPINAPSE: grant agreement No 818399). Further, we acknowledge support from the Swedish Research Council (VR), the Knut and Alice Wallenberg Foundation (KAW), the Foundation for Strategic Research (SSF), Energimyndigheten, eSSENCE and StandUP. We acknowledge help from Dr. Patrik Thunström, with the spin-polarized core calculations.

**Authors contributions:** U.K., A.E., and M.S. conducted the experiments. The experimental data were analyzed by U.K., A.E., M.S., N.H, D.W., and A.A.K. A.A.K. and D.W. designed the experiments. D.I. performed ab initio calculations and A.B. performed the spin-dynamics simulations. M.I.K., L.N. and O.E. provided additional theoretical support on all calculations. All authors contributed to the writing of the manuscript.

**Competing interests:** None declared.

**Data and materials availability:** All data needed to evaluate the conclusions in the paper are present in the paper or the Supplementary Materials. Any additional data pertaining to this manuscript can be made available, pending approval, via the corresponding author of the manuscript.

**Figure captions**

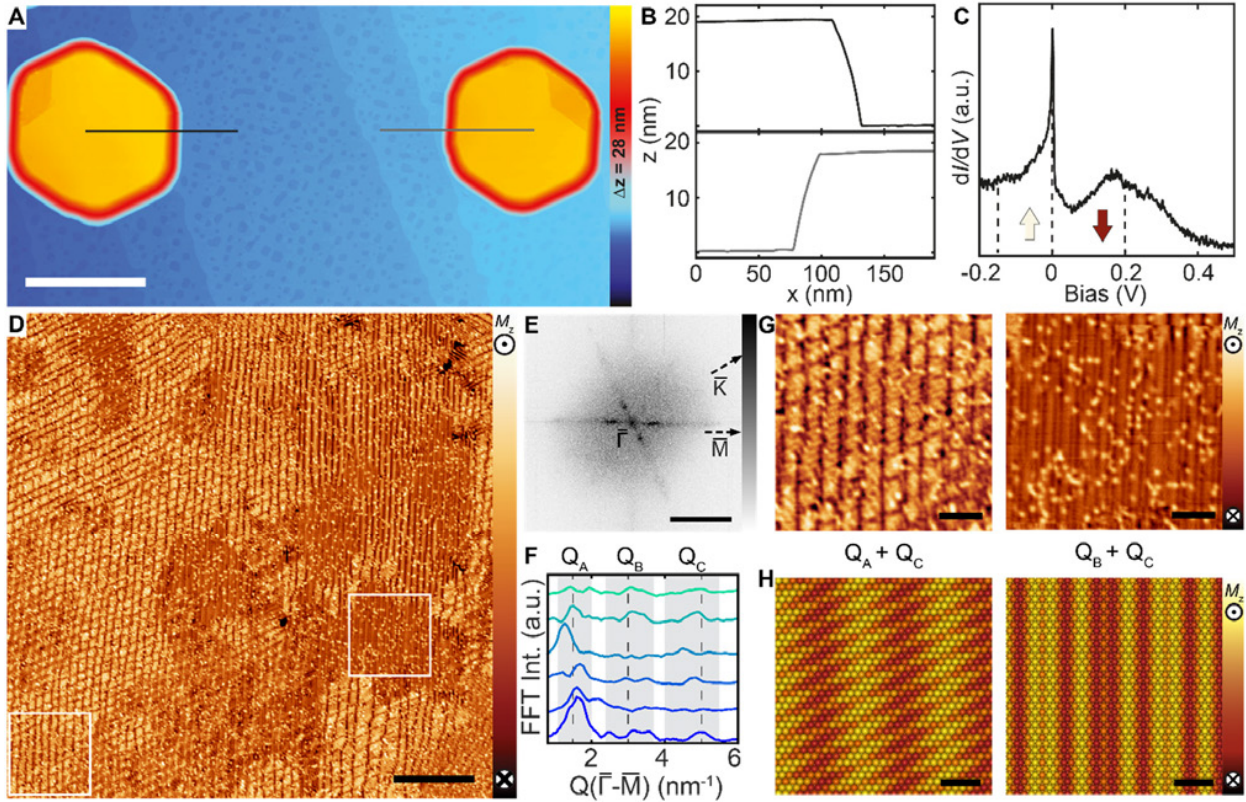

**Fig 1: Imaging the spin Q-glass state of Nd(0001)**: (**A**) Constant-current STM image of a Stranski-Krastanov grown Nd film on W(110), revealing nearly flat-top islands on a Nd wetting layer. The surface defect concentration is less than 0.01 ML (scale bar = 150 nm, $V_S$ = 1 V, $I_t$ = 20 pA). (**B**) Line profiles along the indicated island edges. The thickness of each island is larger than 50 ML. (**C**) d$I$/d$V$ spectrum acquired on the Nd(0001) surface, showing the exchange-split surface state ($V_{stab}$ = 1 V, $I_{stab}$ = 200 pA, $V_{mod}$ = 1 mV; $T$ = 40 mK). (**D**) Magnetization image illustrating the spatially complex magnetic ground state of a spin-Q glass, which lacks long-range order ($T$ = 1.3 K, $B$ = 0 T, $I_t$ = 200 pA). The contrast is directly related to variations in the out-of-plane magnetization ($M_z$) imaged with an out-of-plane sensitive Cr bulk tip (scale bar = 50 nm). (**E**) Q-space image of the magnetization image in (D) (scale bar = 5 nm$^{-1}$), illustrating a large distribution of states in Q space. (**F**) Line-cuts along $\bar{\Gamma}$-$\bar{M}$ of various Q-space images taken from smaller sections of the image in (D) (cf. Fig. S5). (**G**) Close-up views of regions marked in (D), illustrating the local spatial variation of the magnetic order (scale bar = 10 nm). (**H**) Schematic of the out-of-plane projected magnetization resulting from a superposition of the labelled Q vectors (scale bar = 2 nm). The wave vector amplitudes used are marked with dashed lines in (F).

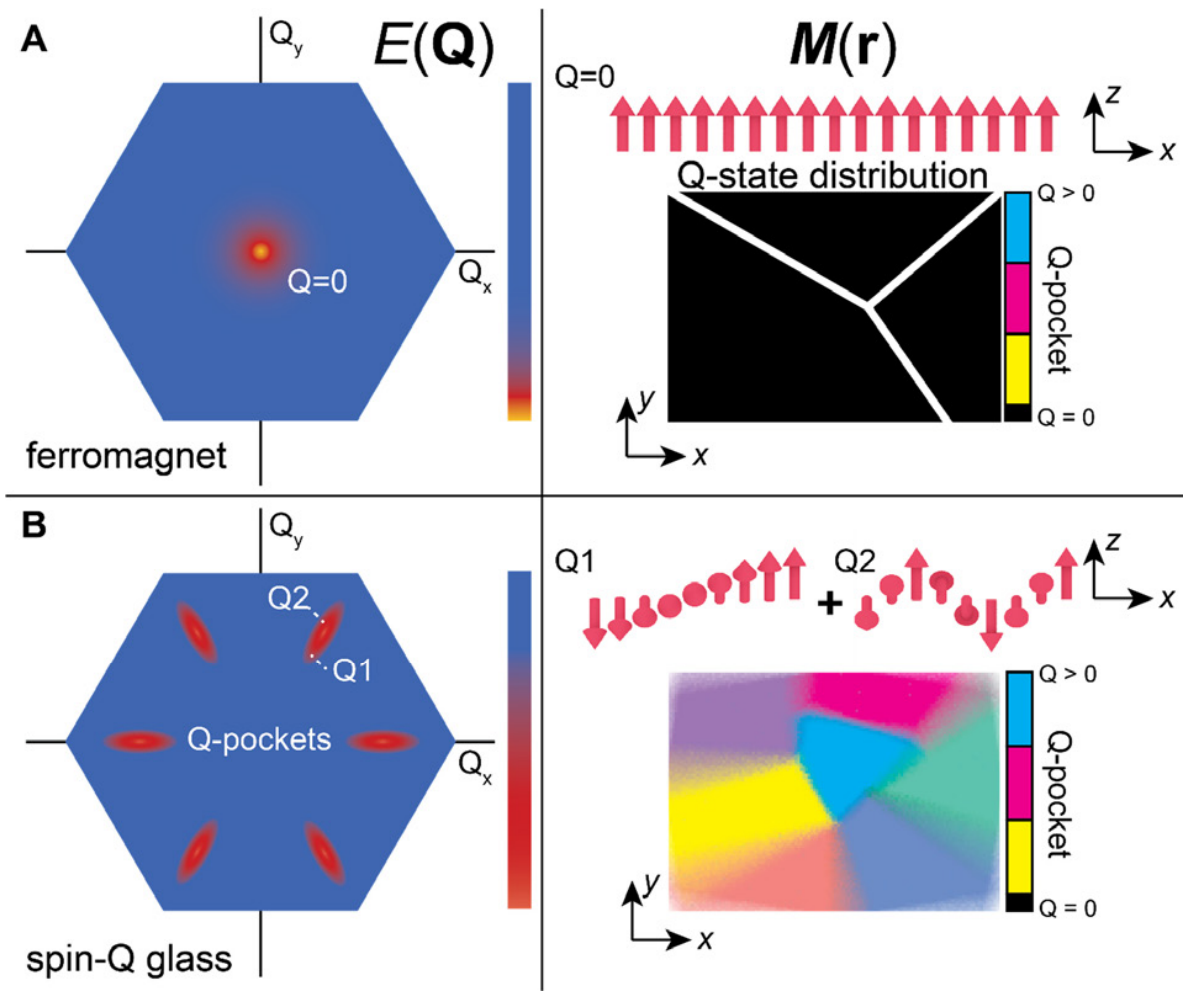

**Fig 2: Energy landscape of a spin-Q glass:** (**A**) Q-space image of the energy landscape $E(\mathbf{Q})$ of a prototypical ferromagnet, which exhibits a strong global minimum at Q = 0, corresponding to a real-space magnetization pattern $\mathbf{M}(\mathbf{r})$ where all spins are aligned. Further minimization leads to the formation of distinct domains (black), separated by domain walls (white), but all domains are defined by a repeating Q-state distribution, where Q = 0. (**B**) A characteristic Q-space image for a spin-Q glass, which can be distinguished by flat valleys (Q-pockets) at non-zero Q values, which leads to a superposition of a distribution of Q states with different periodicities residing in each pocket. This results in a complex $\mathbf{M}(\mathbf{r})$ pattern that lacks long-range order. The spatial distribution of Q states contains regions with local order defined by mixing of Q states (colors) derived from the given Q pockets.

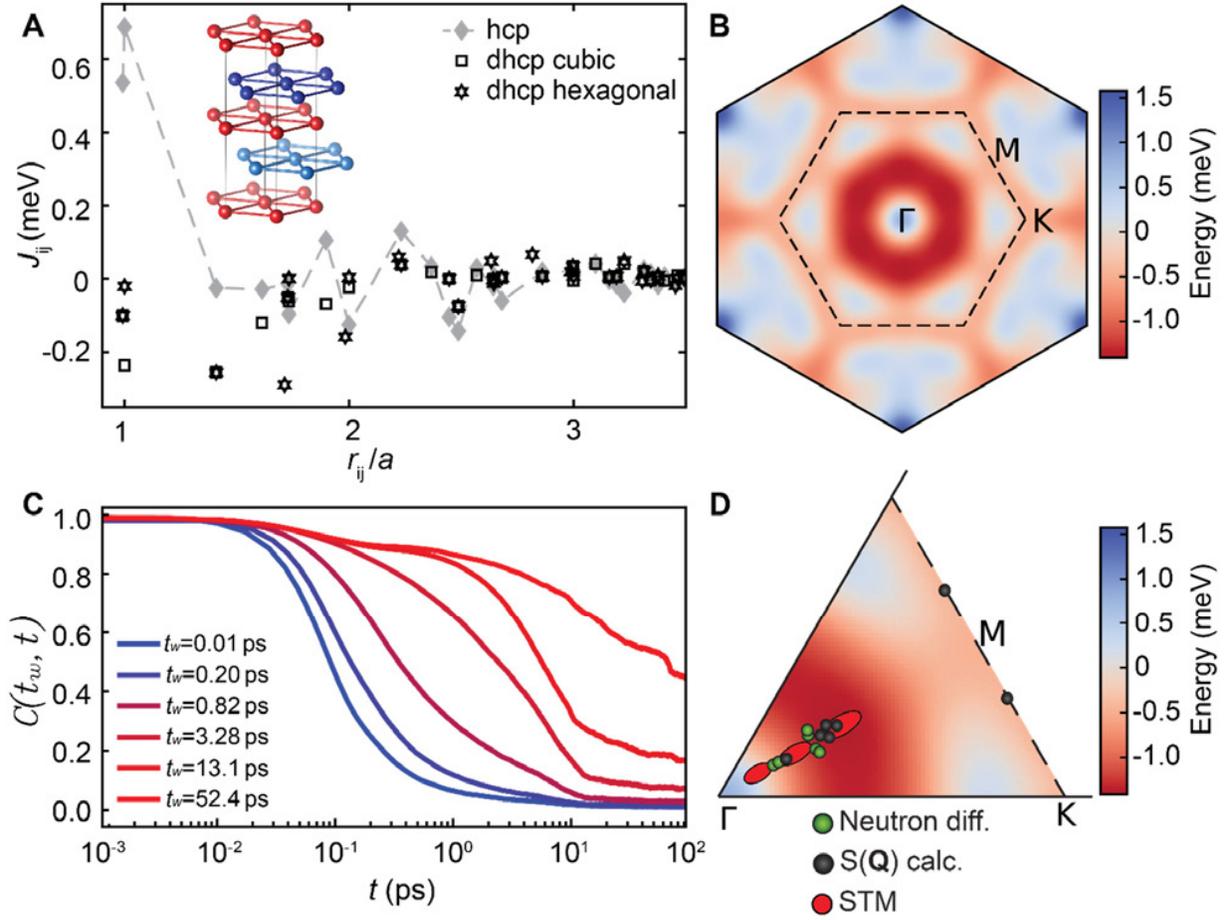

**Fig 3: Elemental Nd electronic and magnetic landscape:** (**A**) Calculated Heisenberg magnetic exchange interactions among Nd spin moments with magnitude 2.454 $\mu_B$, both in the dhcp Nd (black) and the hypothetical hcp structure (gray). A negative interaction denotes a preference for an antiferromagnetic alignment among the spins, while a positive one denotes a ferromagnetic alignment. Inset: the dhcp crystal structure, with an ABAC stacking, where the cubic A sites are represented by red spheres, and the hexagonal B and C sites by light and dark blue spheres, respectively. (**B**) Energy landscape for single-Q spin spirals with $Q = (Q_x, Q_y, 2\pi/c)$ as evaluated from the calculated exchange interactions for the dhcp structure. (**C**) Autocorrelation function $C(t_w, t) = \langle \boldsymbol{m}_i(t + t_w) \cdot \boldsymbol{m}_i(t_w) \rangle$ for dhcp Nd at $T$ = 1 K. (**D**) Comparison of Q states for: SP-STM (this work, red), neutron diffraction (Ref. (*21, 25*), green) and simulations (this work, black).

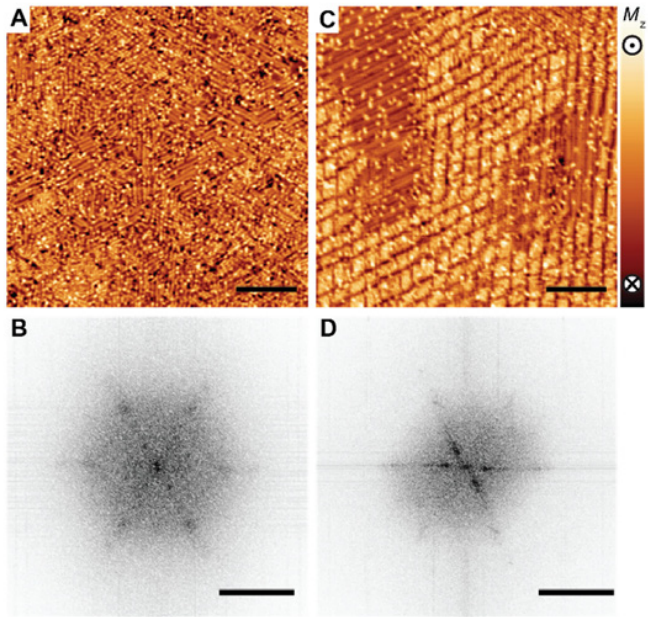

**Fig 4: Effect of defects on the spin-Q glass state:** Magnetization and the corresponding Q-space images of (**A**, **B**) a dirty surface (surface defect concentration of 0.03 ML) and (**C**, **D**) a clean surface (<0.01 ML) (scale bar = 20 nm, $I_t$ = 200 pA for magnetization images, and scale bar = 5 nm$^{-1}$ for Q-space images, inverted grey scale). Higher amount of contamination results in pinning of the Q-state around the defects, as illustrated in real space magnetization images. Overall, measurements of several samples showed that the Q-state distribution is essentially unaltered for concentrations ≤ 0.014 ML. The Q-space image of the dirty sample shows more well-defined Q-pockets as a consequence of the pinning, in contrast to the smeared out Q-pockets along the high symmetry axes in the clean sample.

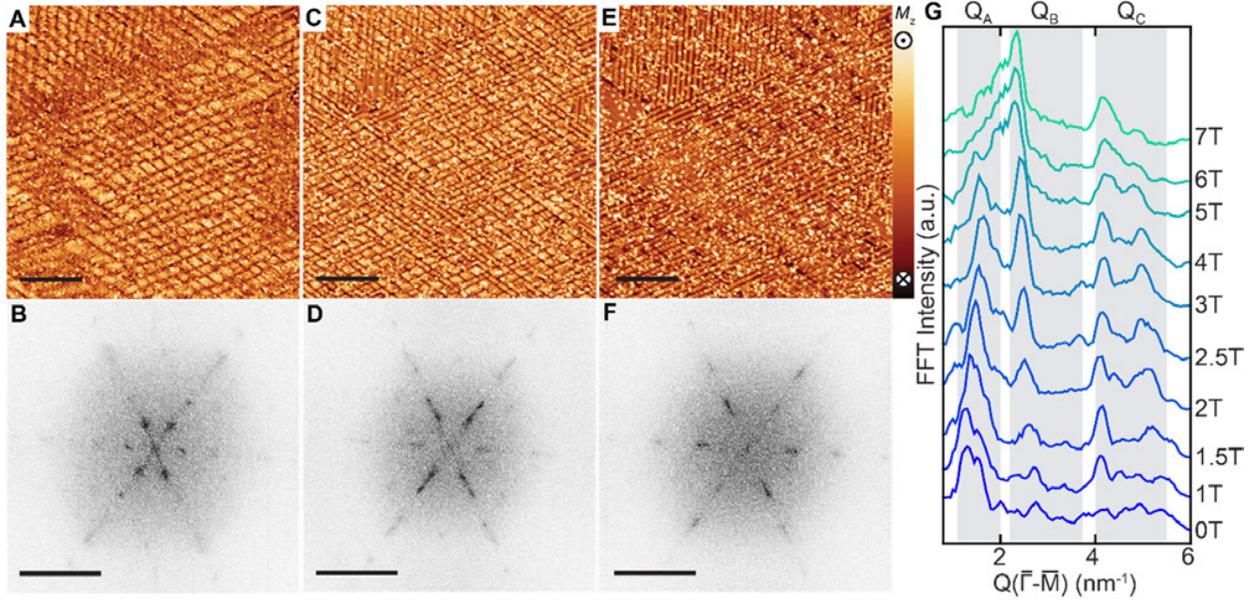

**Fig 5: Magnetic field evolution of the spin-Q glass state:** Magnetization and the corresponding Q-space images of the same area measured at $T$ = 1.3 K, (**A**, **B**) in $B_z$ = 0 T, (**C**, **D**) in $B_z$ = 4 T, and (**E**, **F**) in $B_z$ = 7 T (scale bar = 30 nm, $I_t$ = 200 pA for magnetization images, and scale bar = 4 nm$^{-1}$ for Q-space images, inverted grey scale). Increasing magnetic field does not favor a low Q state, nor exhibits the favorability of any Q state, illuminated by the broadening of the spectral weight in the various pockets. (**G**) Line-cuts along $\bar{\Gamma}$-$\bar{M}$ of Q-space images (Fig. S11) with finer intervals of increasing $B_z$. The spectral distribution becomes smeared out within the given pockets leading to no well-defined long-range periodicity at the highest $B_z$. The surface defect concentration is 0.01 ML.

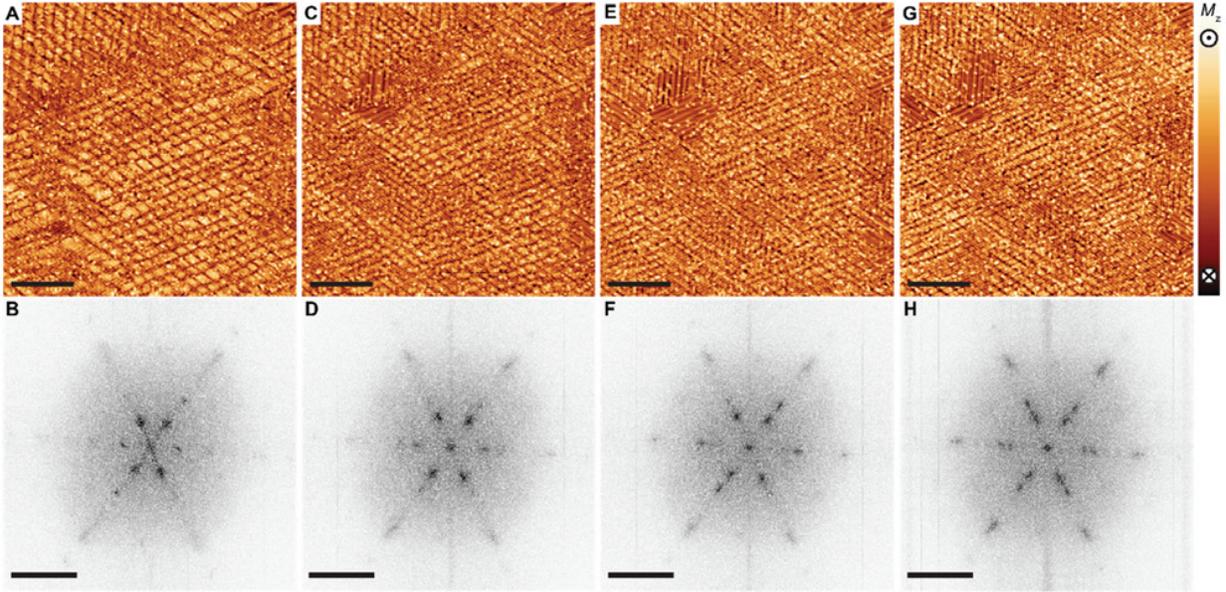

**Fig 6: Aging and glassy behavior of the spin-Q glass state:** (**A**, **B**) Magnetization and the corresponding Q-space images of the surface at $T = 1.3$ K, in its pristine state at $B = 0$ T. The same measurements were performed at the exact same area in $B = 0$ T after subsequent magnetic field sweeps and intermittent probing, leading to the magnetization and Q-space images (**C**, **D**) after $B_{z,1} = +4$ T, $\tau_1 = 10^5$ s, (**E**, **F**) after $B_{z,2} = -4$ T, $\tau_2 = 10^5$ s and (**G**, **H**) after $B_{z,3} = +7$ T, $\tau_3 = 10^5$ s. Typical sweep rates were on the order of 225 mT/min. The sequence shows that the system never reverts to the initial zero-field-cooled state (scale bar = 30 nm, $I_t = 200$ pA for magnetization images, and scale bar = 4 nm$^{-1}$ for Q-space images, inverted grey scale). The surface defect concentration is 0.01 ML.

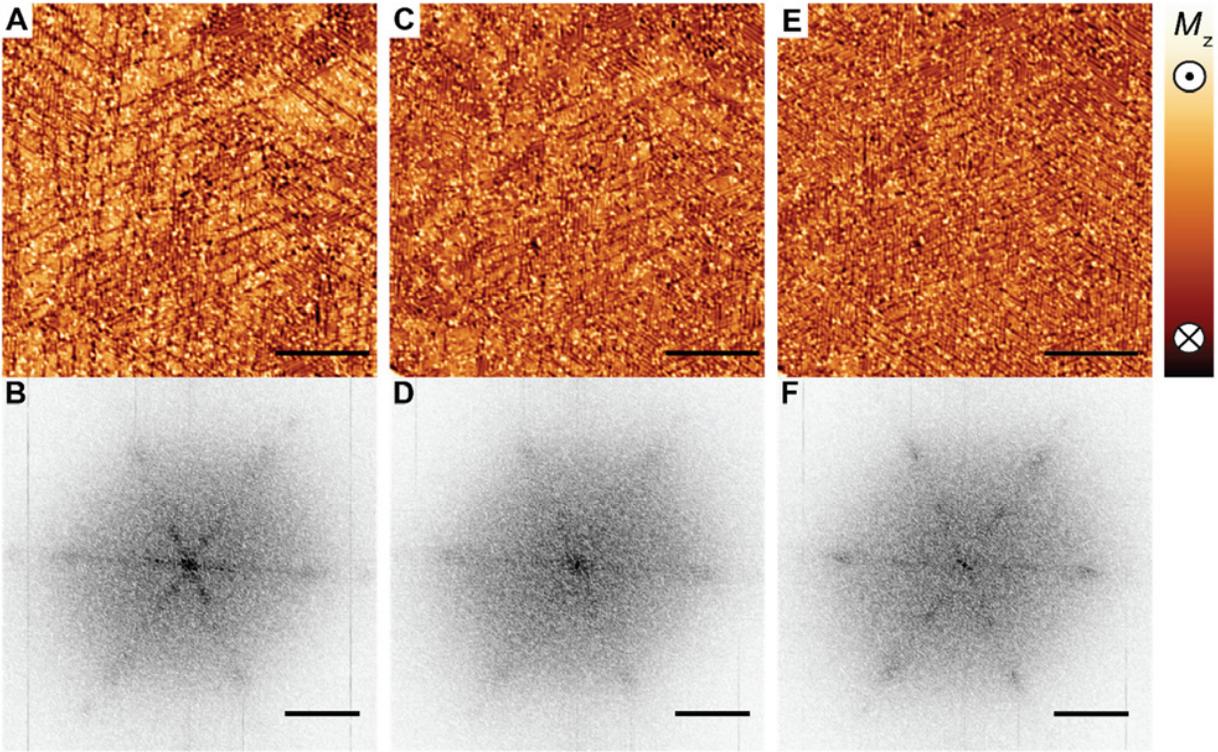

**Fig 7: Temperature dependence of the spin-Q glass state:** Magnetization and the corresponding Q-space images of the exact same are (**A**, **B**) at $T$ = 1.3 K and (**C**, **D**) at $T$ = 4.2 K, in $B$ = 0 T. Warming up the surface from 1.3 K to 4.2 K results in depopulation of the $Q_A$ pocket. (**E**, **F**) The same measurements were performed after +4T magnetic field sweep in the same area at $T$ = 4.2 K in $B$ = 0 T, illustrating similar out-of-plane magnetic field aging behavior (scale bar = 30 nm, $I_t$ = 200 pA for magnetization images, and scale bar = 3 nm$^{-1}$ for Q-space images, inverted grey scale). The surface defect concentration is 0.01 ML.

Supplementary Materials

# Self-induced spin glass state in elemental and crystalline neodymium

Umut Kamber, Anders Bergman, Andreas Eich, Diana Iuşan, Manuel Steinbrecher, Nadine Hauptmann, Lars Nordström, Mikhail I. Katsnelson, Daniel Wegner, Olle Eriksson, Alexander A. Khajetoorians

correspondence to:  a.khajetoorians@science.ru.nl, d.wegner@science.ru.nl

# Contents

## S1 - Growth and morphology of Nd(0001) SK islands and closed films

The motivation for growing films of Nd rather than using bulk single crystals comes from the well-known difficulty of cleaning the surface of bulk crystals of the lanthanide metals, due to severe segregation of bulk impurities toward the surface (*30, 37, 73*). Only after extremely long sputter-anneal cycles, few groups reported successful reduction of bulk impurity segregation such that the surface state could be observed, although surface contamination was still detectable via Auger electron spectroscopy (AES). The alternative approach of epitaxially growing lanthanide films on various 4*d* and 5*d* single crystal substrates (mostly bcc(110) surfaces of W, Mo and Nb were used) has been studied extensively in literature, resulting in high-quality crystals (mosaicities <0.1°) with a (0001) surface orientation and with impurity concentrations below the AES or ARPES detection limit (*29, 31, 35, 36, 39, 41, 44, 45, 48, 74-78*). It was confirmed that various lanthanide (including Nd) thin films of >10 monolayers (ML) are strain-free and possess bulk lattice constants; the electronic structure is fully developed at 30 ML, and long-range magnetic order is bulk-like in the range of 30-50 ML (*27, 29, 31-33, 35, 36, 38-49, 51, 74-93*). For Nd, the dhcp structure with bulk lattice constants was confirmed, independent of the substrate used (*34, 35, 38, 46, 77*). Moreover, neutron and XRD scattering studies on films and superlattices found magnetic peaks identical to those of bulk samples for thin Nd films down to 33-39 ML (*38, 46, 47*). We note that in one study of Nd/Sm superlattices additional magnetic satellite peaks have been observed that are not known from any Nd bulk studies (*77*), however, as Sm also orders magnetically, this can be ascribed to some interlayer coupling effects of the superlattice. In summary, epitaxial lanthanide films of sufficient thickness provide superior bulk and surface cleanliness while fully maintaining all bulk properties.

Measurements were performed on epitaxial islands of Nd(0001) grown on W(110). The W(110) substrate was cleaned by repeated cycles of annealing at $T$ = 1250°C in a gradually reduced oxygen atmosphere (from $p$ = 1 x 10$^{-7}$ mbar in the first cycle down to $p$ = 2 x 10$^{-8}$) and flashing at $T$ = 2400°C, as described in Ref. (*94*). W(110) surfaces that showed insignificant contamination of oxygen or carbon were used for subsequent Nd growth, as done previously in temperature-dependent experiments (*27, 54*). The Nd source material was purchased from AMES laboratory (www.ameslab.org; purity 4N, see Table S1 for the most abundant elemental impurity concentrations), which was melted and then thoroughly degassed under UHV conditions inside the crucible of an electron-beam evaporator, in order to further reduce the contamination from oxygen and hydrogen. For the growth of Nd films, the material was sublimated from the electron-beam evaporator and deposited onto the W(110) kept at room temperature. Subsequent annealing at $T$ = 700°C for 15 minutes resulted in either a Stranski-Krastanov (SK) growth when the nominal film thickness was smaller, or an ordered closed film with an atomically smooth surface (*48*). Fig. S1A illustrates a typical large-scale STM image after growth of ca. 15 ML Nd and subsequent annealing. Large islands of Nd, on the order of hundreds of nanometers in diameter and > 50 ML high were typically observed (cf. line profile in Fig. 1B). The surface was relatively flat with atomic steps visible from either locally varying thickness or overgrown steps at the interface to the W substrate (Fig. S1B). Such islands showed the expected exchange-split surface state of Nd(0001) (Fig. 1C) (*27, 54*). Fig. S1C shows the topography of a ~100 ML thick Nd film after the same annealing procedure. The larger nominal coverage would have required a much higher annealing temperature to lead to SK growth (*48*). In this case, the annealing temperature was sufficient to create a well-ordered and atomically smooth surface, but the layers did not yet break up to form the thermodynamically more stable SK islands. A major difference in the morphology of the two samples was the presence of dislocations. Despite an overall high crystallinity, extended dislocations could be identified in the image area of the closed thick film (*51, 80*), while all sufficiently annealed SK islands were free of dislocations.

At this point, we would like to comment on the morphology of the sample. AC-susceptibility (*42, 48*), SP-STM (*49, 82, 94*) and STS (*27, 49, 54, 79, 92*) studies on various lanthanide metals showed that SK-grown islands reflect bulk magnetic properties, provided they are sufficiently thick. The advantage of SK-grown islands is that strain relief is better than in layer-by-layer grown films, that is misfit dislocations (due to lattice mismatch at the interface) and screw dislocations (caused by antiphase boundaries created in the initial growth) can be avoided (*39, 40, 49, 91*). To check whether there might be any finite-size effect present in our samples, we investigated different islands with various volumes that is with heights ranging from 58 to 92 ML and with lateral areas ranging from 58000 to 200000 nm$^2$. All results reported here were consistently observed on all these differently sized islands. Moreover, we also studied the 100 ML-thick closed film to further verify that our islands were not in the thin-film limit and did not show any lateral finite-size effects. As will be discussed in the next section, the closed film indeed displayed the same behavior as the islands mostly studied. Our preference of focusing on islands was based on two observations: (a) the closed films exhibited much dirtier surfaces (Fig. S3D), as more deposited material simply leads to more impurities that can segregate to the surface; (b) the screw dislocations found in the closed film, but not for the islands, can pin the magnetic structure, as was demonstrated for thick Dy films on W(110) (*51, 80*). Hence, SK-grown islands have a superior morphology and crystallinity, and from our comparison of differently sized islands and the closed film we could rule out any finite-size effects. Thus, we conclude that the SK-grown nanocrystallite islands fully reflect the structural, electronic and magnetic properties of bulk neodymium.

**S2 - Real-space magnetization and corresponding Q-space images**

Experiments were performed in two different home-built UHV-STM systems. The first system can be operated at two different stable temperatures, $T$ = 1.2 K and $T$ = 4.2 K, with an applied out-of-plane magnetic field up to 9 T. The second system operates at $T$ = 30 mK and $T$ = 7 K, with an in-plane magnetic field up to 4 T, and an out-of-plane magnetic field up to 9 T (*69*). Our systems allow the tip to stay on the same sample area while switching between the mentioned temperatures. As spin-polarized measurements require a stable drift-free tip-sample tunneling contact, we are neither able to acquire data continuously while warming up or cooling down to the stable temperature regimes, nor can we stabilize the tunneling junction at temperatures above the previously mentioned temperatures. After a magnetic field sweep or temperature change, we need to stabilize the tunneling junction on the order of 60 min, before resuming an image on an identical area. Similarly, spatially dependent imaging cannot be performed during a magnetic field sweep due to similar stabilization arguments resulting from small changes in temperature, combined with the length of time it takes to acquire a typical image (120 min). Thus, aging experiments only contain images before and after magnetic field exposure, with varying exposure times.

Fig. S2A illustrates the surface of a typical Nd island, imaged with an out-of-plane polarized bulk Cr tip at $V_S$ = 1 V. Note that for all STM topography images shown in the main text as well as the supplementary material, apart from a global plane subtraction no processing of the raw data was performed. The surface of these islands showed a distribution of surface impurities, which likely resulted from oxygen, carbon or hydrogen contamination (*52, 53*), in addition to features that were related to the underlying morphology of the W(110): substrate step edges were visible on the Nd surface because the atomic layer height for the two materials is different (*52*). Tunneling spectroscopy of the surface (Fig. 1C and Fig. S2I) revealed the expected surface state of Nd(0001), which is characterized by an exchange splitting into a majority peak visible below $E_F$ and a minority peak above $E_F$, with an additional narrow peak at $E_F$ (*54*). Previous studies have measured the temperature dependence of the exchange splitting $\Delta E_{\text{ex}}$, illustrating that the splitting persists far above $T_N$ = 19.9 K (*27*). The slowly diminishing value of $\Delta E_{\text{ex}}$ above $T_N$ is unusual compared to other lanthanide metals (*27, 49, 81*), which is another signature of strong local correlations in Nd (*26*). We note that the narrow peak at $E_F$ has been seen on other lanthanides as well (*27, 53, 54*), i.e., there is no connection to the spin-Q glass behavior only found in Nd so far. Comparing spectra at $T$ = 1.3 K and $T$ = 40 mK, this resonance narrowed at lower temperature. Also, we did not see this feature strongly change in magnetic field. Therefore, we believe this feature is most likely related to a van Hove singularity. We note that images taken at a sample bias of 1 V (Fig. S2A) did not show any spin contrast. This is expected, as tunneling then occurs not only into the minority surface state but also bulk majority and minority states. Hence, the STM image is identical to that taken with a nonmagnetic tip.

In order to acquire spin contrast, we utilized a Cr bulk tip, which we cleaned by *in-situ* electron bombardment, and later prepared by tip pulsing until achieving an optimal out-of-plane contrast. We note that we could sometimes observe probes with a canted magnetization, but we have never observed probes with only a pure in-plane and stable magnetization. Zochowski *et al.* already pointed out that there is a tendency toward a simplified separation of spin order on the hexagonal vs. the cubic sites (*23*), whereas in fact: "It has been established that, below $T$ = 10 K, the hexagonal sites have induced moments with components along the *c*-axis due to the ordered moments on the cubic sites and that the cubic sites have induced moments with components in the *c*-axis due to the ordered moments on the hexagonal sites" (*23*). This statement is further supported by thermal expansion and magnetostriction data showing anomalies along the *c*-axis in the entire temperature range from 1 to 10 K. Hence, we expect that an out-of-plane polarized tip can map the out-of-plane component of the spin structure, irrespective of whether the surface is terminated by a hexagonal or a cubic layer. As STM probes the topmost layer, we have no means of determining what the termination of our surfaces is. Nevertheless, we point out that we measured islands of various thicknesses, sometimes including monolayer step edges. The magnetic structure was always observed to be comparable.

Constant-current SP-STM images near either peak of the exchange-split surface state using out-of-plane tips showed the magnetic pattern in real space that is overlaid with the topographic features (Fig S2B, C). While the topographic contrast was nearly identical at both imaging biases, the magnetic contrast was nearly fully inverted. This is a hallmark of spin polarization. To verify this, we also compared tunneling spectra from two different local points on the surface that showed opposing magnetic contrast (Fig. S2I). As expected, the red spectrum had a larger intensity of the majority peak below $E_F$ compared to the blue spectrum, whereas the situation was reversed for the minority peak above $E_F$ (*80, 82*). Tunneling into bulk states is negligible in the energy window governed by the surface state, because there is a bulk symmetry gap in the center of the surface-projected Brillouin zone. Thus, tunneling almost exclusively happens between the tip and the highly spin-polarized surface state, i.e., out of the majority state at negative sample bias and into the minority state at positive sample bias, respectively (*94, 95*). As a further evidence that the observed contrast was of magnetic origin, we present in Fig. S2J a constant-current STM image of an island acquired with a non-magnetic W tip, using the same stabilization parameters as in Fig. S2B (note that this image is from a different sample preparation where the surface impurity density was higher). No magnetic contrast could be observed with the W tip, as opposed to the Cr tip. The W tip could, however, be made spin-polarized by dipping the tip into Nd and thereby picking up a Nd cluster at the tip apex. Fig. S2K shows the exact same area as in (J) using such a Nd-terminated tip, which now showed the magnetic contrast of the surface. Due to the lack of knowledge about both the polarization and the behavior of the tip in magnetic field, we chose to use well-known bulk Cr tips for the rest of this study.

For the magnetization images introduced in the main text, we acquired SP-STM images at $V_S$ = 200 mV and $V_S$ = -150 mV, respectively (Fig S2B, C), and subsequently subtracted them from each other (Fig S2D). As a result of the subtraction process, the real-space magnetization images did not require any further processing (e.g. background correction or filtering), since both of the topography images have the same global slope. As the topographic contrast in both images was nearly identical, and the magnetic contrast was inverted, subtraction resulted in the magnetic contrast without the large-scale topographic features. This also revealed that that overgrown steps from the underlying W(110) interface did not affect the magnetic structure on the surface, as we did not observe any disturbance in the magnetic pattern correlating with the substrate step edges. We note that most atomic-scale impurities are still visible after subtraction. However, this is an electronic effect rather than magnetic contrast, because the impurities influence the width and intensity of the much narrower majority state and the van Hove singularity more strongly than the much broader minority state, which leads to a bias-dependent apparent height of the impurities such that they do not cancel in the subtraction (*53*). For the Q-space images, we computed the Fast-Fourier Transform (FFT) of the real space images by using MATLAB, and we did not apply any post-processing, e.g. symmetrization or removal of intensity near $Q = 0$ that is caused by edge effects of the image, discussed in detail below (cf. section S3). The FFT intensity is always shown with square root scaling in inverted grayscale. All the Q-space images shown in this study have minimum 12 pixel per nm$^{-1}$ resolution.

By comparing the Q-space images for the individual SP-STM images taken at each surface state energy (Fig. S2F,G) with the Q-space image of to the subtracted magnetization image (Fig, S2H), we illustrate that the features stemming from the magnetic structure were nearly identical. This is also true for images acquired in magnetic field (cf. Fig. S12). We note here that tips with a magnetization component in the plane of the sample led to certain observed *Q* vectors in a given image at one surface-state energy which was different compared to its surface-state partner image (see circles in Fig. S2).

In Fig. S3, we compare the topography, magnetization and Q-space images of a Nd island (A-C) with those of the ~100 ML closed film (D-F). Despite the fact that the closed film had a higher surface-defect density (D, see also Fig. 4 and the corresponding discussion in the main text) and a few screw dislocations, the magnetization image (E) clearly showed a multi-*Q* structure with locally varying short-range order but no globally ordered ground state, reminiscent of the image observed on the island (B). This was also reflected by the corresponding Q-space image (F), which qualitatively agreed well with that of the island (C). This confirms our conclusion of the previous section that SK-grown islands, despite their seemingly small size, fully reflect bulk magnetic properties.

**S3 - Variation in local order and Q-state distribution**

To determine the range of Q-pockets mentioned in the main text ($Q_A$, $Q_B$, and $Q_C$), we considered several Q-space images of the subtracted real-space magnetization images, measured at $T = 1.3$ K on several islands of similar dimensions without an external magnetic field. Various line-cuts along the $\overline{\Gamma}$-$\overline{M}$ direction of the Q-space images were produced in order to visualize the spectral weight of the *Q* vectors. Line-cut data (Fig. 1F, 5G, S13I) were flattened by subtracting a linear baseline, and smoothed by utilizing a Savitzky-Golay filter with 7-9 neighbors in order to reduce the background noise.

In Fig. S4, we utilized inverse FFT filtering at each identified Q-pocket and plot the resultant filtered magnetization images, where the overall intensity is related to the contribution of the selected *Q* states.

The images illustrate the large variation in spectral weight of each of the Q-pockets and that the underlying local order has various contributions from the named Q-pockets as schematically illustrated in Fig. 2B of the main text. In Fig. 1F of the main text, we show line-cuts of FFTs taken from smaller regions of the large-scale image illustrated in Fig. 1D. Each of these regions, plotted in Fig. S5, primarily showed one type of distribution of $Q$ states.

Here, we would like to comment on the resolution of the Q-space images in relation to the observed widths of features. In brief, the resolution of a 2D FFT depends on a number of factors, e.g. the number of pixels in the image compared to the wavelength of the features extracted. As the FFT is performed always on finite-size images with a finite number of pixels, there are artifacts, well known in the STM community that can arise. For example, the enhanced intensity near $Q = 0$ in the FFT is a result of finite-size effects, and this intensity must be disregarded. Usually, this is cropped out of images, but we preferred to provide unprocessed raw data. Most importantly, the interpretation of reciprocal peak widths as known from neutron diffraction or XRD, cannot be translated one-to-one to Q-space maps, extracted from SP-STM. Spot broadening in diffraction experiments is commonly interpreted as an effect of finite sample or domain size. To exemplify that our Q-space maps have to be interpreted differently, we point out the examples shown in Fig. 1 of the main text: the smaller images in Fig. 1G and Fig. S5 were simply cropped from the larger image (Fig. 1D). The FFT of the larger image, i.e. larger spatial area, showed a much broader and smeared out intensity (Fig. 1E), compared to the cropped images (Fig. S5D-F, J-L). The cropped images showed much more well-defined Q-vectors, but with varying magnitudes and spectral intensities. In essence, if one were to "sum" these smaller regions up, a broadened intensity would be obtained, which is characteristic of the spin-Q glass landscape we highlight in the paper. That said, the discussed broad features were real and not due to limited resolution of our FFTs. This can also be deduced from the magnetic field-dependent data where features sharpened along the high-symmetry direction (angular direction), while the magnitude of the Q-vectors smeared out (radial direction). The FFTs simply reflect what is seen in the real space images, namely that the wavelengths are not uniform. Magnetization images showed a spatially dependent spectral weight of $Q$ states, as illustrated in the FFTs of Fig. 1E of the main text and Fig. S5.

### S4 - Comparison of surface and bulk magnetic moments

STM is a surface-sensitive method, and therefore care has to be taken before one can conclude that it reflects the bulk magnetization. In order for the surface layer(s) of a magnetic material to possess a different magnetic structure than the bulk, it is first of all necessary that the magnetic state (as reflected by moments and exchange interactions) is drastically different at the surface layers. Secondly, these exchange interactions need to be sufficiently dominant in order to overcome the exchange coupling between atoms at the surface and atoms deeper into the bulk. In order to investigate the difference in magnetism of the bulk and the surface of Nd we performed calculations, as described in the methods section, of two surface terminations of the Nd crystal, with either a cubic layer as the surface layer or the hexagonal layer as the surface layer. The former calculation employed a slab geometry with 13 atomic layers and the latter calculation employed a slab with 11 layers. In each calculation we introduced a vacuum region of 11 Å. The structures were relaxed with respect to geometry, using force minimization, and it was found that the surface layer of the hexagonal termination relaxed 2.5% inward, while the subsurface layer relaxed 6% outward. For the cubic termination the surface layer relaxed 2.2% inward while the subsurface layer relaxed 5.8% outward.

The magnetic moments of the two surface calculations are shown in Table S2. Here we report the total moment of the surface (1$^{st}$) and subsurface (2$^{nd}$) layer. These moments should be compared to those of the bulk, which are also presented in Table S2. It may be seen that the total moment was similar for the surface layers (of either termination) and the bulk values. In addition, we found that the bulk value of the total magnetic moment was reached 3-4 atomic layers below the surface (data not shown). The microscopic reason why the magnetic state is bulk-like just a few layers below the surface can be traced back to the electronic structure, that in essentially all materials investigated by density functional theory, it is known to become bulk-like just a few atomic layers below the surface (*96*). This has been discussed in terms of the 'nearsightedness' of density functional theory (*97*).

**S5 - Spin simulations and Q-state energetics**

The spin simulations presented in the main text were performed using both Monte Carlo simulations (MC) and atomistic spin dynamics simulations (ASD), using the UppASD software (*71*). We used the parameterized spin Hamiltonian $H = -\sum_{i \neq j} J_{ij} \boldsymbol{e}_i \cdot \boldsymbol{e}_j$ based on unit vector spins $\boldsymbol{e}_i$ and exchange couplings $J_{ij}$ calculated from DFT and the RSPt software (*70*), as described in the methods section. The ground state of the two methods was the same, and for equilibrium properties the methods can be used interchangeably. The ASD methodology gives a correct description of the timescale for the dynamical processes in the system and thus a possibility to control the rates of relaxation processes. For the MC simulations we used the Metropolis-Hastings algorithm (*98*) to obtain states as close to the ground state as possible. Given the glassy nature of the system, the efficiency of this procedure depended strongly on the relaxation protocol, i.e. how the temperature was varied during the simulations.

The ASD simulations were performed by evaluating the Landau-Lifshitz-Gilbert equation of motion for each atomic moment $\boldsymbol{m}_i$ in the effective site-dependent field $\boldsymbol{B}_i$ according to

$$\frac{d\boldsymbol{m}_i}{dt} = -\gamma \boldsymbol{m}_i \times \boldsymbol{B}_i - \gamma \frac{\alpha}{m_i} \boldsymbol{m}_i \times (\boldsymbol{m}_i \times \boldsymbol{B}_i)$$

where $\gamma$ is the gyromagnetic ratio, and $\alpha$ the Gilbert damping parameter that determines the rate of dissipation of energy from the spin system. Thermal effects were included in the ASD method by means of Langevin dynamics where a stochastic noise term is added to the effective magnetic field $B_i$. All simulations were done with the UppASD software (*71, 99*) and the algorithms used are given in full detail in Ref. (*72*).

For the determination of Q states from the static correlation functions $S(Q)$, displayed in Fig. 3D in the main text, we followed a simulated cooling protocol where the temperature was decreased from $T$ = 20 K (which is above the ordering temperature) with steps of $\Delta T$ = 5 K while performing $10^5$ Monte Carlo sweeps at each temperature. The correlation functions were then sampled using $10^5$ ASD steps with a Gilbert damping parameter of $\alpha = 0.01$ and a time step of $10^{-16}$ s. The Q-space correlation function $S(Q)$ was obtained as the Fourier transform of the real-space trajectories of the system, which were sampled every $100^{th}$ time step.

The energetics of the single-Q spirals, displayed in Fig. 3B in the main text, were evaluated by constructing helical spin spirals for given wave vectors $Q$, where the plane of rotation of the spins was defined to be perpendicular to the wave vector, i.e. the magnetization of the spin-spiral has both in-plane and out-of-plane components. The spatial magnetization profile for a single-Q spin spiral was constructed as:

$$\boldsymbol{m}(\boldsymbol{r}, \boldsymbol{Q}) = m\,[\,\sin(\boldsymbol{r} \cdot \boldsymbol{Q})\,\hat{\boldsymbol{n}} + \cos(\boldsymbol{r} \cdot \boldsymbol{Q})\,\hat{\boldsymbol{z}}]$$

where $\hat{n} = \hat{Q} \times \hat{z}$ so that the in-plane component of the magnetization is perpendicular to the $Q$ vector. The energy of each single-Q spiral was then obtained by evaluating the spin Hamiltonian defined above. We note that since only isotropic Heisenberg exchange interactions were included in the Hamiltonian, the energy of a single-Q spin spiral did not depend on the orientation of the plane where the spins rotate, i.e., cycloidal and helical spin-states were degenerate for the same $Q$ vector.

**S6 - Periodicity of the spin-spiral energy landscape**

The single-Q spin spiral energy landscape $E(Q)$, which is presented in Fig. 3B in the main text, has a periodicity that might seem confusing: when going along the Σ symmetry line (Γ-M-Γ), it would seem that $E(Q)$ is not equivalent for the two Γ points. This behavior can, however, be explained from the dhcp structure, in particular by the difference between the positions of the atoms on the hexagonal and the cubic sublattices. In the Figs. S6-7, we plot schematically how the spins of the different sublattices rotated in the presence of a wave vector $Q$. For simplicity, we considered a ferromagnetic reference state but the analysis holds for the ground state magnetic structure of bulk dhcp Nd, where the moments of the different sublattices are rotated 90 degrees with respect to each other as well. In the schematic figures, the moments of the two inequivalent sites for the cubic sublattices were colored with different shades of red, while the moments on the hexagonal sublattices were colored blue, i.e., using the same color representation as the crystal structure displayed in Fig. 3A in the main text.

### S7 - Decomposition of the simulated magnetic structure for various *Q* vectors

The schematic magnetization images shown in Fig. 1H in the main text were obtained as linear combinations of spin spirals with distinct *Q* vectors having different wavelengths ($Q_A$ = 1.5 nm$^{-1}$, $Q_B$ = 3.0 nm$^{-1}$, and $Q_C$ = 5.0 nm$^{-1}$), inspired by the observed Q-pockets. This is analogous to an inverse Fourier transform of the selected *Q* vectors, with the difference that for the construction of the spirals used for the images in Fig. 1H, the magnetization of the resulting multi-*Q* spin spirals was rescaled so that the norm of the magnetization was kept constant. A proper linear combination of single-*Q* spirals would otherwise result in longitudinal fluctuations, i.e. changes of the local magnetic moment magnitudes; but in order to stay consistent with our unit-vector based Hamiltonian, we here enforced constant moment magnitudes. In Fig. S8, we illustrate the resultant site-dependent magnetization for various *Q* vectors taken from the calculation illustrated in Fig. 3 in the main text.

As a supplement to the schematic picture in Fig. 2B in the main text, we simulated a quenching, i.e. a rapid cooling of the system going from *T* = 7 K to *T* = 0 K, using Metropolis MC simulations (Fig. S9). The quenching protocol ensured a correct description of the long-range magnetic structure of the high-temperature state while removing local fluctuations. From this rapidly cooled system several single-*Q* and multi-*Q* states were visible, where the spectral weight varied spatially without clearly defined domain walls as in typical multi-*Q* systems (*58*). These results were comparable to the experimental observations in Figs. 1 and 5.

### S8 - Calculation of the autocorrelation function

The results for the autocorrelation function $C(t_w, t) = \langle m_i(t + t_w) \cdot m_i(t_w) \rangle$ presented in the main text were obtained from ASD simulations following the Landau-Lifshitz-Gilbert equation described in preceding sections. The procedure was to start with a completely random distribution of magnetic moments, whose directions evolved with time. For a selection of waiting times $t_w$, separate single-site correlations $m_i(t + t_w) \cdot m_i(t_w)$ were sampled during the simulation times, *t*, and averaging was performed over all Nd sites. Here a large value of the Gilbert damping parameter $\alpha = 0.5$ was used in order to capture the relaxation dynamics efficiently. While such a large damping can be seen as unrealistic, earlier ASD simulations on spin glass systems have shown that the glassy relaxation dynamics is captured for all considered choices of damping values (*10*). For comparison, autocorrelation measurements were also performed for simulations with lower damping values as well as when using the Metropolis MC algorithm, and all simulations showed glassy behavior of the autocorrelation functions.

In order to further establish the glassy behavior of the dhcp Nd system, and to also confirm that the short time scales available from the atomistic simulations are still long enough to identify different types of relaxation dynamics, additional simulations on dhcp model systems were performed. In addition to the proper bulk dhcp Nd system, we here considered model spin Hamiltonians containing interactions from only the nearest, next-nearest, and third-nearest neighbors. For these short-ranged Hamiltonians we then modified the interactions to be either (a) with the same magnitude of the dhcp Nd interactions but with positive sign to all interactions, giving a ferromagnetic ground state; (b) same magnitude of the dhcp Nd interactions but with negative signs to all interactions, giving a frustrated/antiferromagnetic ground state; or (c) random exchange interactions following a Gaussian distribution, renormalized to mimic the same order of magnitude of the exchange interactions of dhcp Nd, giving an Edwards-Anderson like model (*1, 2*). The dynamics of all model Hamiltonians were simulated using ASD with a Gilbert damping factor of 0.5 and for system sizes consisting of a 24x24x24 repetition of dhcp cells. We note that this differs from the 32x32x32 size used for the autocorrelation simulations presented in the main text.

The simulations, displayed in Fig. S10, showed that for the ferromagnetic model Hamiltonian, the autocorrelation function converged to a finite value rapidly as expected with the large Gilbert damping used. The antiferromagnetic model also showed a rapid convergence of the autocorrelation function. In contrast, the Edwards-Anderson model showed a relaxation behavior closer to that of the spin-Q glass behavior of dhcp Nd presented in Fig. 3C in the main text. It is also evident from Fig. S10 that there was a distinct difference of the autocorrelation functions for the glassy systems (spin-Q and Edwards-Anderson) compared to the non-glassy (ferro- and antiferromagnetic) systems which was clearly visible for the simulation times available from the ASD simulations.

**S9 - Magnetic field dependence of spin-Q glass**

Fig. S11 illustrates all the magnetic as well as Q-space images in varying out-of-plane magnetic field that were utilized for the line-cuts displayed in Fig. 5G of the main text. With increasing out-of-plane magnetic field, we observed that the $Q_A$ and $Q_B$ pockets merged, while the spectral weight more broadly distributed in the $Q_C$ pockets. These features were independent of the subtraction method, as shown in Fig. S12, where we compared the raw SP-STM images and their respective FFTs at the two surface-state energies, for two different magnetic fields. Comparing Figs. S12C,D with Fig. S11F as well as Figs. S12G,H with Fig. S11Q, each image showed the same features as the corresponding subtracted image. Similar trends were seen for in-plane magnetic fields.

Fig. S13 illustrates a region measured at $T$ = 40 mK, with a higher defect density (ca. 0.015 ML). The overall spectral weight collapsed onto the axis perpendicular to the applied field direction, which is indicated in the figure. The same qualitative behavior has also been observed in field-dependent neutron diffraction studies with magnetic fields applied close to one of the basal high-symmetry directions (*22-24*). At intermediate fields, a variation in spectral weight was observed, similar to the out-of-plane experiments. This spectral weight broadened in higher applied fields, as illustrated by the line cuts shown in Fig. S13I. We note here that due to the high defect density, the FFTs of this region were noisier in comparison to the out-of-plane data.

**S10 - Aging in Nd(0001)**

The distinguishing feature of spin glasses is the presence of aging dynamics. The most traditional definition of aging, as inspired by experimental evidence, is characterized by magnetization states which never fully relax resulting from the presence of multiple relaxation time scales, regardless of the waiting time (cf. section S8). This relaxation behavior can span many orders of magnitude in time, as exemplified by the magnetic alloys such as Cu-Mn (*3, 100*), which exhibits glassy dynamics below a particular freezing temperature above which the material is paramagnetic.

Aging behavior in out-of-plane magnetic fields is extensively discussed in the text. We also saw aging behavior for applied in-plane magnetic fields. Fig. S14 illustrates a region measured at $T$ = 40 mK. We note that in the pristine state (Fig. S14A), there were predominately Q-states along one orientation. As this area was imaged near an edge of the island, we suggest that magnetostrictive strain might have locked the $Q$ states into this direction during the initial zero-field cool-down of the sample. However, after applying an in-plane field sweep, the $Q$ states completely randomized their orientation (Fig. S14B), uninfluenced by the island edge. This illustrated the negligible impact of the island edges to the energy landscape. After subsequent in-plane field sweeps, we saw similar behavior as in out-of-plane fields, in which the spectral weight was randomly changed and showed no preferential distribution. Moreover, to compare the behavior with the above data taken at $T$ = 1.3 K, we also performed a field sweep in an out-of-plane field at $T$ = 40 mK, on the same area (Fig. S14E), and observed the same aging behavior that is illustrated in the main text. Therefore, we conclude that the aging behavior was independent of the applied field direction.

## S11 - Temperature dependence of spin-Q glass state

Fig. S15 illustrates the magnetic field dependence and resultant aging at $T$ = 4.2 K. Comparison of magnetization images taken at $T$ = 1.3 K and $T$ = 4.2 K, as well as aging behavior for the Q-pockets are discussed in detail in the main text. Imaging the same region at $T$ = 4.2 K in out-of-plane magnetic fields also illustrates the same trends in the spectral weight in other regions of Q-space similar to the observations at $T$ = 1.3 K. Namely, the spectral weight broadened along the other pockets without the appearance of any favorable $Q$ state.

     Previous neutron diffraction data illustrates the emergence of new $Q$ states with decreasing temperature, starting at $T$ < 19.9 K, down to $T$ = 1.8 K. Similarly, we performed temperature-dependent measurements. At $T$ = 7 K (Fig. S16A, C), we observed primarily $Q$ states at high $Q$ values with an absence of longer-range ordered states as observed at $T$ = 1.3 K. Images of the same region at $T$ = 40 mK (Fig. S16B, D) illustrated a behavior similar to that observed at $T$ = 1.3 K, with the emergence of spectral weight in the $Q_A$ pockets which are absent in images at $T$ = 7 K. Moreover, we note that we did not observe any new $Q$ states at $T$ = 40 mK, in comparison to images taken at $T$ = 1.3 K, suggesting that either the system was completely frozen into a finite number of $Q$ states, or that fluctuations persisted down to lower temperature, inhibiting certain $Q$ states to freeze.

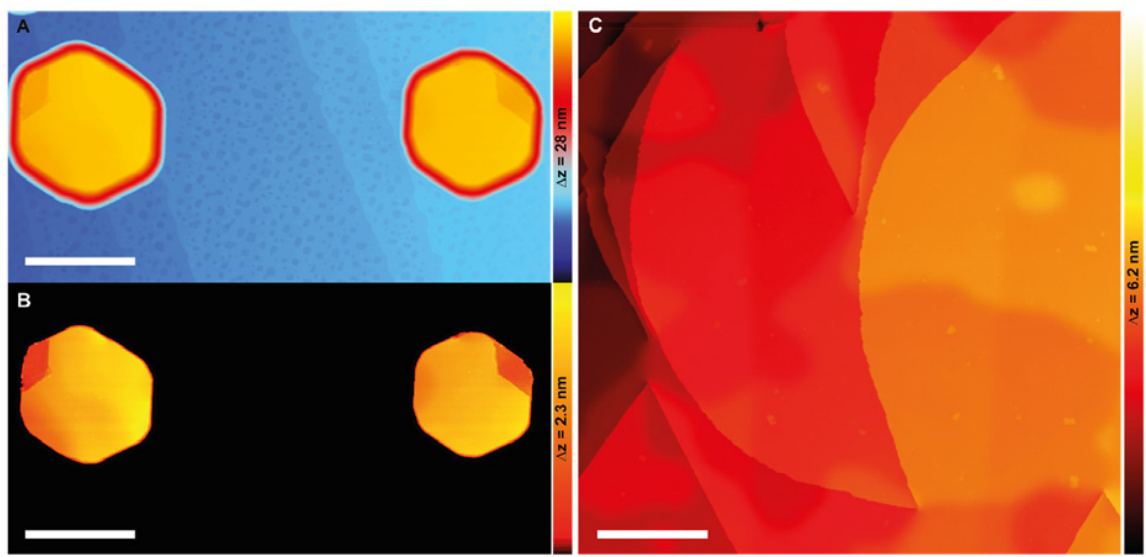

**Fig. S1. Nd(0001) morphology of islands and closed films.**

(**A**) Constant-current STM image of a Nd film on W(110) (~15 ML deposition), after annealing to 700°C, leading to Stranski-Krastanov growth of nearly flat-top islands (> 50 ML) on a Nd wetting layer ($V_S$ = 1 V, $I_t$ = 20 pA, scale bar = 150 nm). (**B**) Same image as in (A) with reduced vertical scale bar to show the atomic-scale surface morphology. Steps are visible due to locally varying Nd layer thickness as well as overgrown atomic steps at the Nd-W interface. No line defects or any other signature of dislocations are visible. (**C**) Constant-current STM image of a ~100 ML Nd film on W(110) after annealing to 700°C, leading to an extended, closed Nd film with relatively homogeneous thickness distribution ($V_S$ = 1 V, $I_t$ = 20 pA, scale bar = 150 nm). Several screw dislocations can be identified.

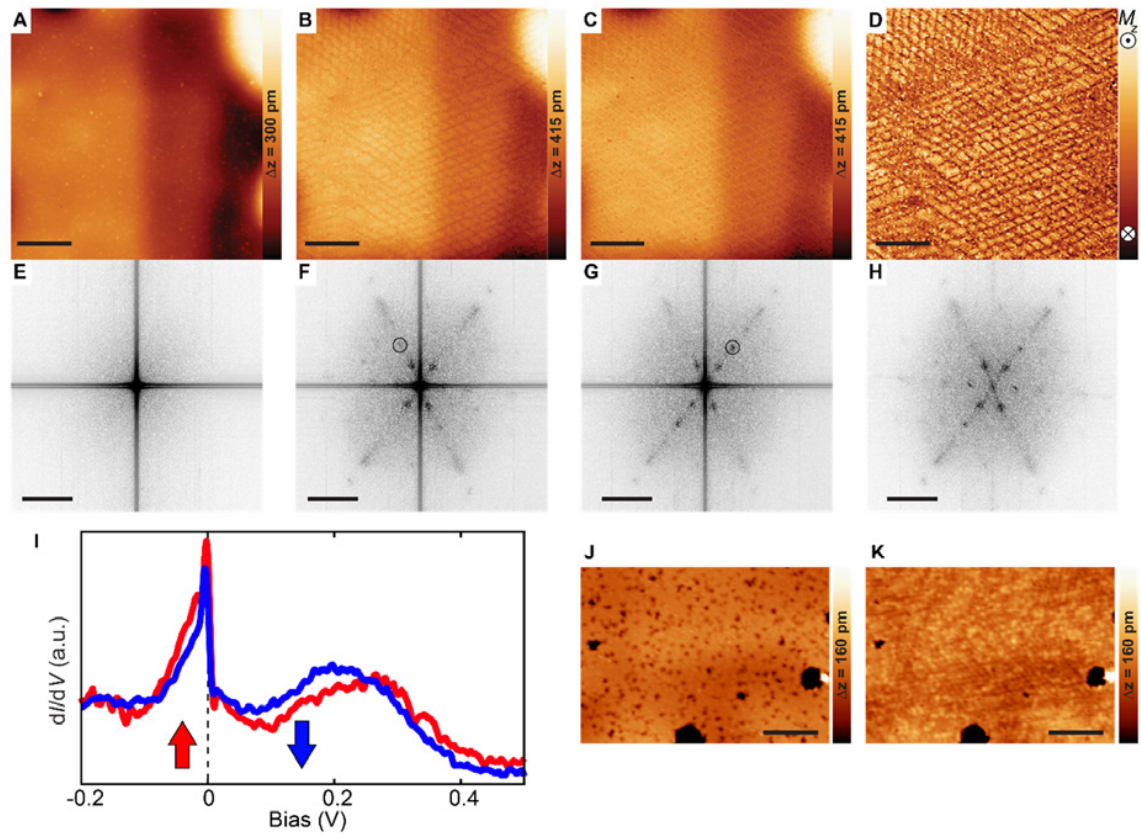

**Fig. S2. Magnetization imaging method**

(**A-C**) Topography images at constant-current mode with various sample biases; (A) at $V_S$ = 1 V, $I_t$ = 200 pA showing the morphology of the island surface; (B) at $V_S$ = 200 mV, $I_t$ = 200 pA and (C) at $V_S$ = -150 mV, $I_t$ = 200 pA both containing also magnetic information (scale bar = 30 nm). Magnetic contrast in the images measured at the minority (B) and the majority (C) surface state exhibits contrast reversal. (**D**) Magnetization image obtained by subtracting the minority image (B) from the majority image (C), revealing the magnetic structure of Nd(0001) surface. (**E-H**) Q-space images of the real-space images in (A-D) obtained by FFT (scale bar = 3 nm$^{-1}$). (**I**) d$I$/d$V$ spectra acquired on two different points, showing opposite magnetic contrast on the surface ($V_{stab}$ = 1 V, $I_{stab}$ = 300 pA, $V_{mod}$ = 2 mV; $T$ = 1.3 K). (**J**) Constant-current STM image of a surface region probed with a non-magnetic W tip at $V_S$ = 200 mV, $I_t$ = 200 pA (scale bar = 15 nm). (**K**) The exact same area as in (J) imaged after picking up a magnetic Nd cluster by dipping the W tip gently into another Nd island, illustrating the magnetic patterns in the real space on top of the topographic features ($V_S$ = 200 mV, $I_t$ = 200 pA, scale bar = 15 nm).

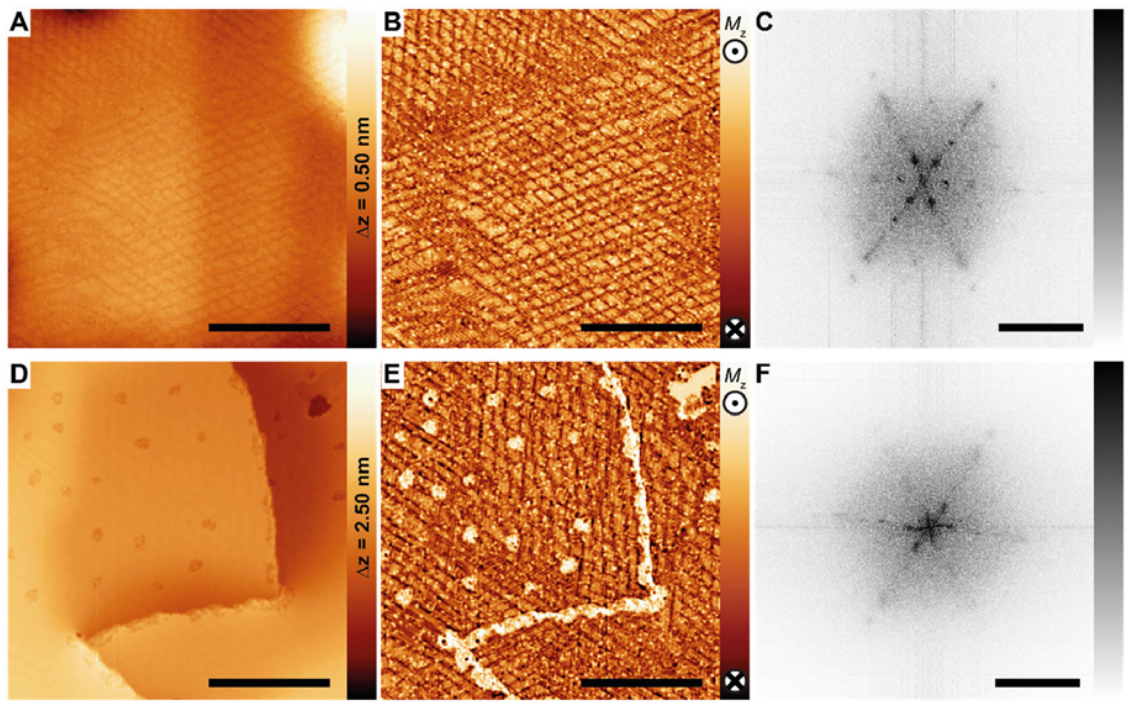

**Fig. S3. Influence of magnetic structure on sample morphology.**
(**A**) Topography of the surface of the Nd island shown in Fig. S2, which has an average thickness of 60 ML ($V_S$ = 200 mV, $I_t$ = 20 pA, scale bar = 50 nm). (**B**) Magnetization image of the same area shown in (A). (**C**) Corresponding Q-space image of the magnetization image in (B) (scale bar = 5 nm$^{-1}$). (**D**) Topography of the surface of a closed ~100 ML thick Nd film. Note the five times larger z-range compared to (A) ($V_S$ = 200 mV, $I_t$ = 20 pA, scale bar = 50 nm). (**E**) Magnetization image of the same area shown in (D). (**F**) Corresponding Q-space image (scale bar = 5 nm$^{-1}$) of the magnetization image in (E).

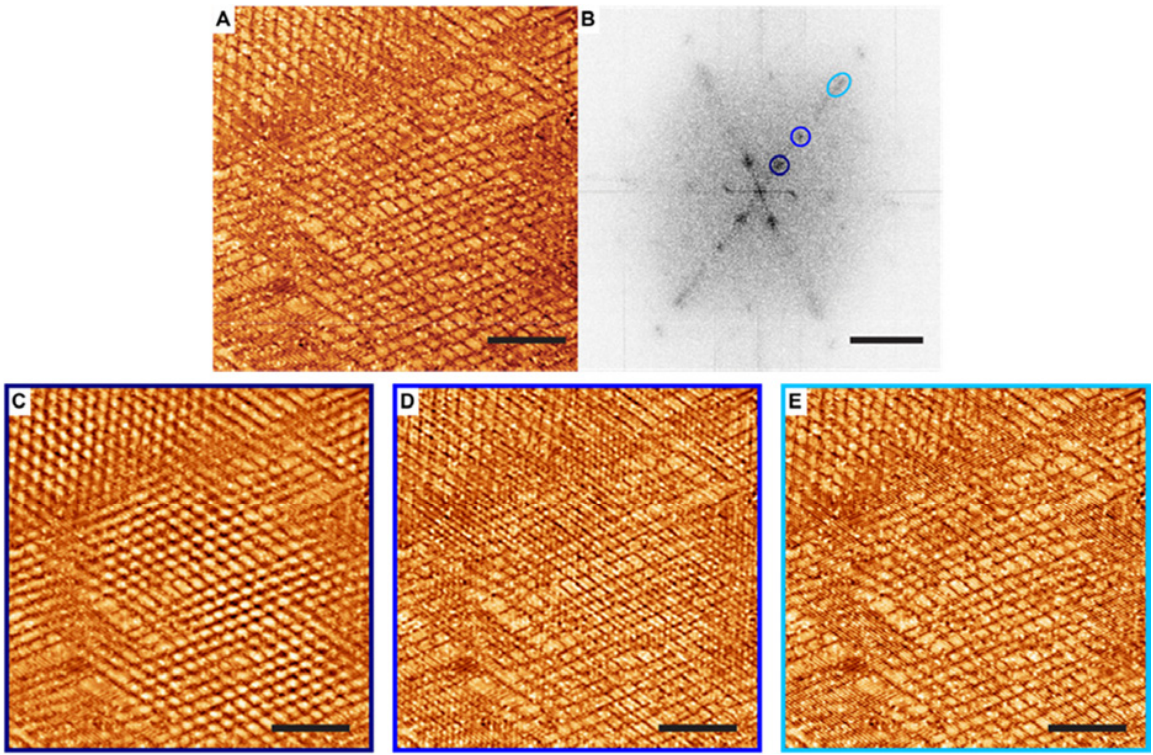

**Fig. S4. Spectral decomposition of Q-pockets in real space.**

(**A**) A magnetization image of the Nd(0001) surface, showing the spatially complex magnetic structure with superposition of different Q-states in different regions (scale bar = 30 nm, $I_t$ = 200 pA). (**B**) Q-space image of the magnetization image in (A) (scale bar = 3 nm$^{-1}$). (**C-E**) Spectral decomposition maps obtained by superimposing the Q-states marked in (B) on the magnetization image in (A). By this way, real-space structure and spatial distribution of each Q-pocket can be visualized, revealing that not only the locations that are present but also the relative intensities are different.

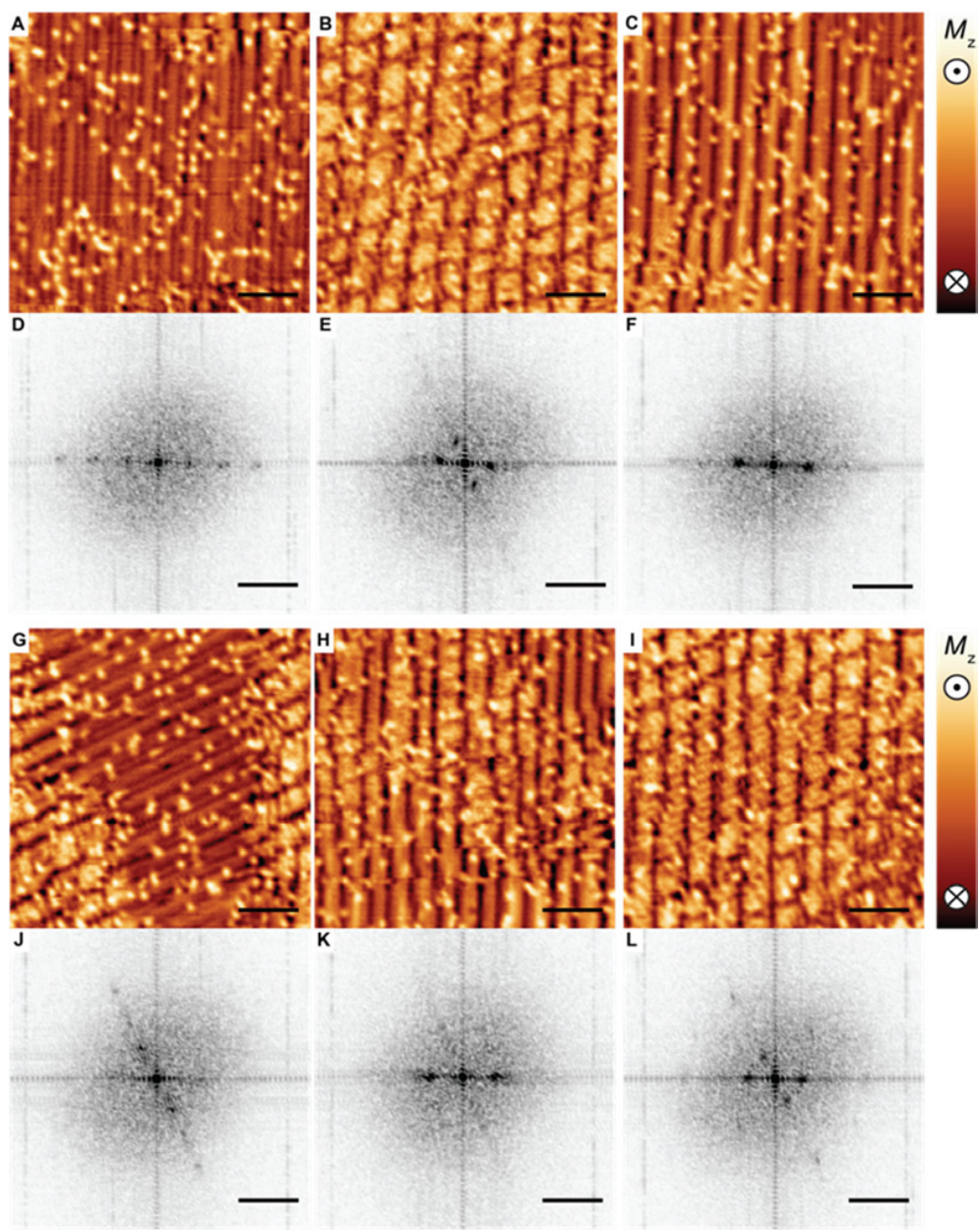

**Fig. S5. Locally ordered states.**
(**A-C**, **G-I**) Magnetization images showing different spatial regions of Fig.1D in the main text (scale bar = 10 nm, $I_t$ = 200 pA). (**D-F**, **J-L**) Corresponding Q-space images revealing the local order, characterized by superposition of $Q$ states (scale bar = 3 nm$^{-1}$).

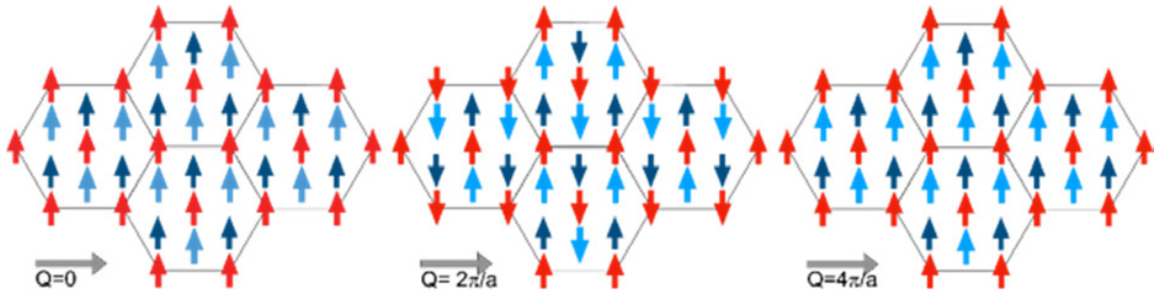

**Fig. S6. Brillouin zone Γ-K.**

Spin rotations for *Q* vectors along the direction of the grey arrow. Colors refer to the sublattice affiliations of the moments (red = hexagonal, blue = cubic). For $Q = 0$ (Γ, left panel), all moments are aligned ferromagnetically. A spiral with $Q = 2\pi/a$ (middle panel) results in a collinear antiferromagnetic state, while increasing the wave vector to $Q = 4\pi/a$ gives the original ferromagnetic state (right panel). For wave vectors along this direction, the periodicity is thus $4\pi/a$.

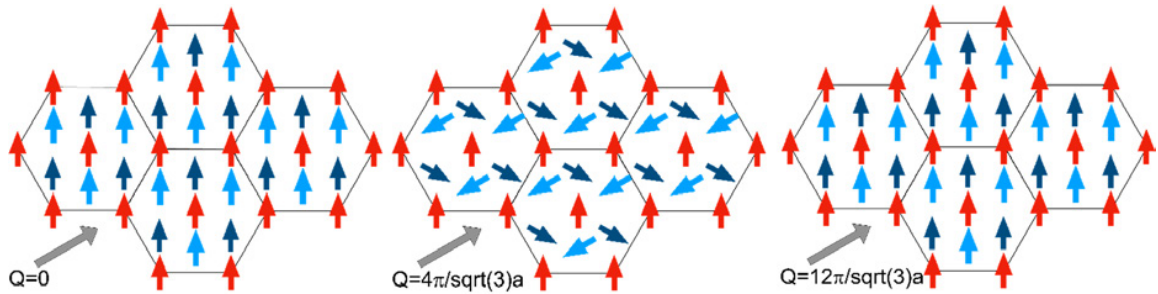

**Fig. S7. Brillouin zone Γ-M.**

Spin rotations for $Q$ vectors along the direction of the grey arrow. For $Q = 0$ (Γ, left panel), all moments are aligned ferromagnetically. A spiral with $Q = 4\pi/\sqrt{3}a$ (middle panel) results in a non-collinear antiferromagnetic state where the moments of each hexagonal sublattice are rotated by 120 degrees with respect to the other sublattices, and all moments on the cubic sites are ferromagnetically aligned. In order to obtain the original ferromagnetic state for all sublattices, the wave vector needs to be increased to $Q = 12\pi/\sqrt{3}a$ (right panel). For wave vectors along this direction, the periodicity is thus longer for the dhcp structure than for the hcp structure.

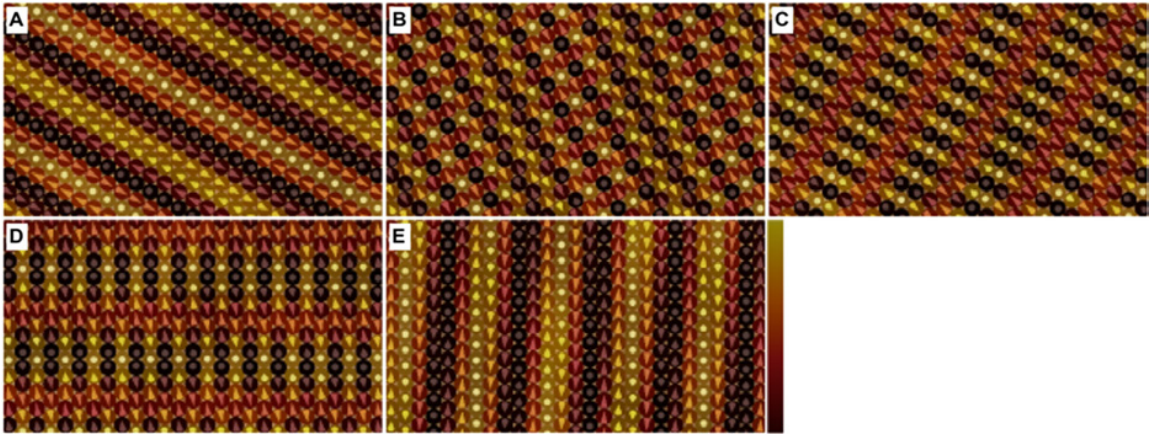

**Fig. S8. *Q* vector decomposition of the calculated real-space magnetization.**

Schematic depiction of the spin structure, where the directions of the cones refer to the magnetization direction, and the color contrast refers to the projected out-of-plane magnetization (colored spheres) obtained as the inverse Fourier transforms of the most significant *S*(*Q*) signals from the ASD/MC simulations. All structures are characterized by a single *Q* vector which is (**A**) *Q* = 2π/*a* [0.22, -0.13, 0.00]; (**B**) *Q* = 2π/*a* [0.24, -0.40, 0.00]; (**C**) *Q* = 2π/*a* [0.44, -0.39, 0.00]; (**D**) *Q* = 2π/*a* [-0.10, -0.85, 0.00]; and (**E**) *Q* = 2π/*a* [0.02, -0.22, 0.00], where *a* is the lattice parameter for dhcp Nd.

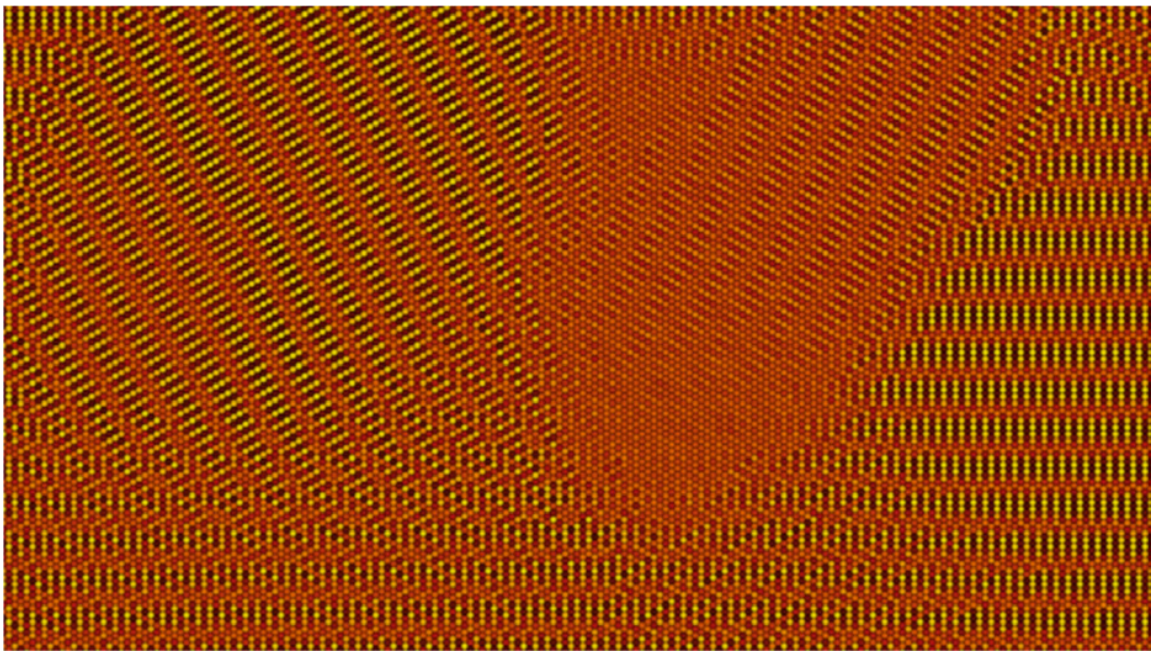

**Fig. S9. Monte Carlo simulated results of a rapid cooling from 7 K to 0 K.**
Results from spin dynamics simulations of bulk Nd as parameterized by *ab initio* exchange interactions, using Metropolis MC simulations.

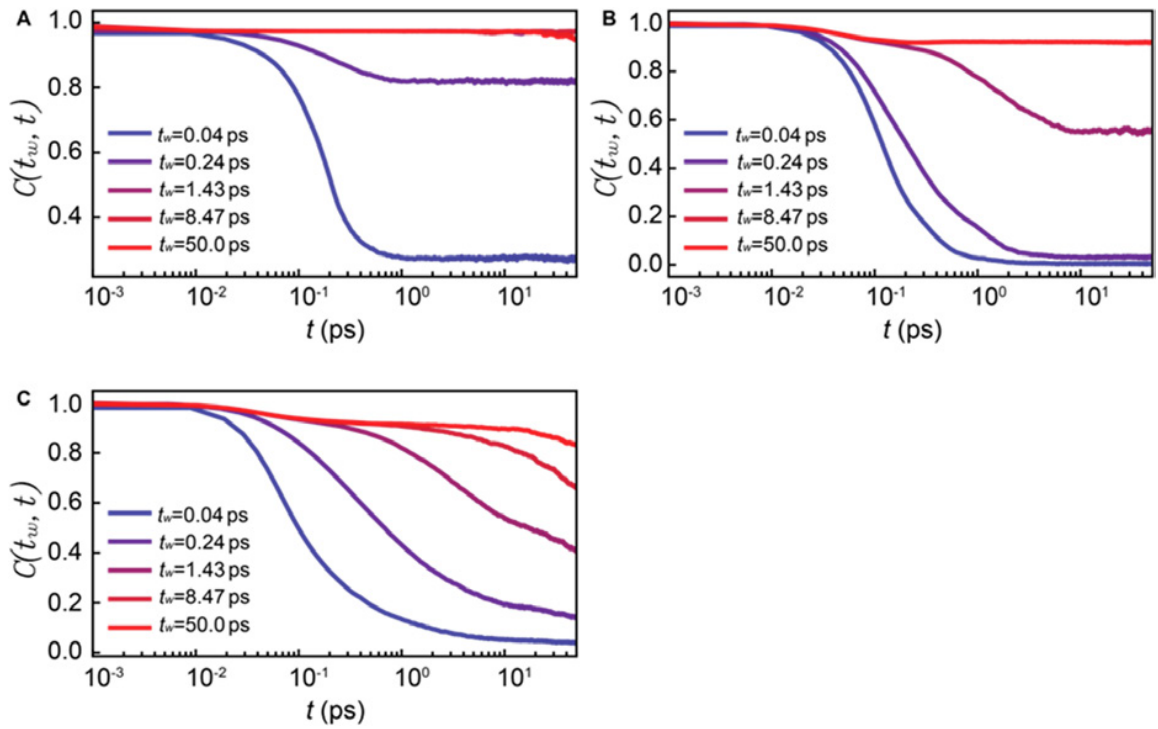

**Fig. S10. Autocorrelation and aging.**
Autocorrelation behavior showing the simulated relaxation dynamics at $T = 1$ K for (**A**) the ferromagnetic model, (**B**) the antiferromagnetic model, and (**C**) the Edwards-Anderson like model.

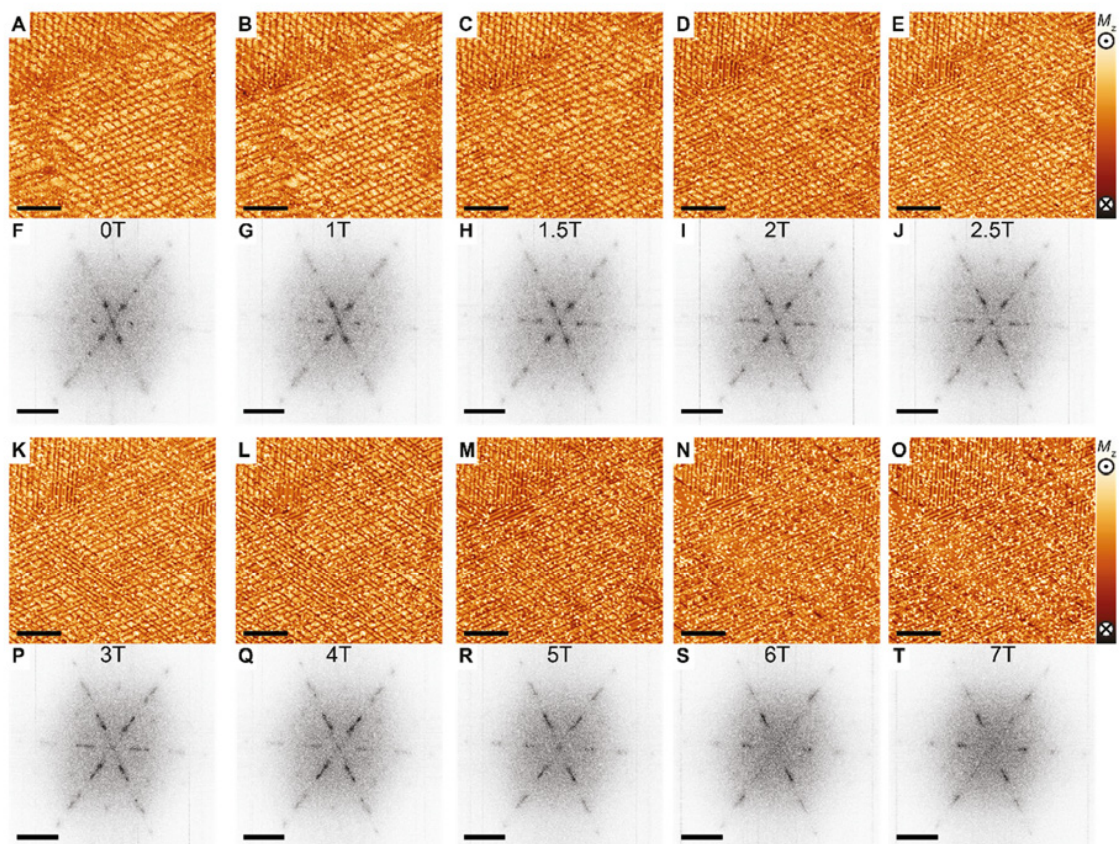

**Fig. S11. Out-of-plane magnetic field evolution of the spin-Q glass state.**
(**A-E**, **K-O**) Magnetization images of the same area measured in various out-of-plane magnetic fields at $T$ = 1.3 K (scale bar = 30 nm, $I_t$ = 200 pA). (**F-J**, **P-T**) Corresponding Q-space images of the magnetization images (scale bar = 3 nm$^{-1}$). All images are plotted with the same contrast.

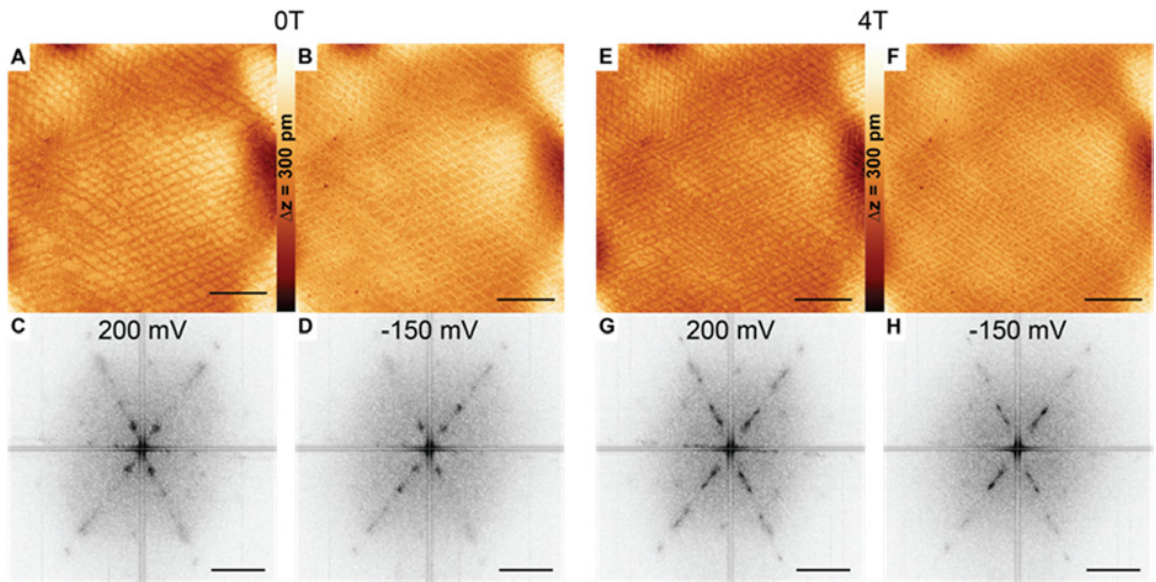

**Fig. S12. Magnetic field dependence of spin contrast at surface-state energies.**
(**A**, **B**) Constant-current SP-STM images measured at $V_S$ = 200 mV, $I_t$ = 200 pA and $V_S$ = -150 mV, $I_t$ = 200 pA at $T$ = 1.3 K in $B$ = 0 T. (**E**, **F**) The same area imaged in $B_z$ = 4 T, at the same two bias voltages (scale bar = 30 nm). (**C**, **D**, **G**, **H**) Corresponding Q-space images of the images above, which can be compared to Fig. S11F and Fig. S11Q, respectively (scale bar = 3 nm$^{-1}$).

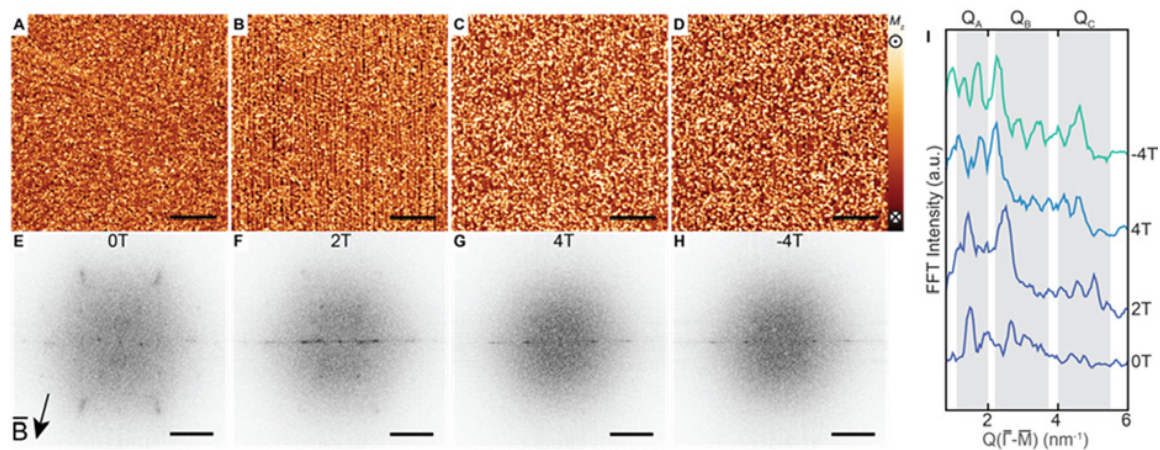

**Fig. S13. In-plane magnetic field evolution of the spin-Q glass state.**
(**A-D**) Magnetization images of the same area in various in-plane magnetic fields at $T$ = 40 mK (scale bar = 30 nm, $I_t$ = 200 pA) and (**E-H**) corresponding Q-space images of each image above (scale bar = 3 nm$^{-1}$). In-plane magnetic field is rotated 15° with respect to the y-axis of the STM images (cf. arrow in (E)). The sign of the magnetic field value defines the direction of the field. (**I**) Line-cuts along $\bar{\Gamma}$-$\bar{M}$ of Q-space images.

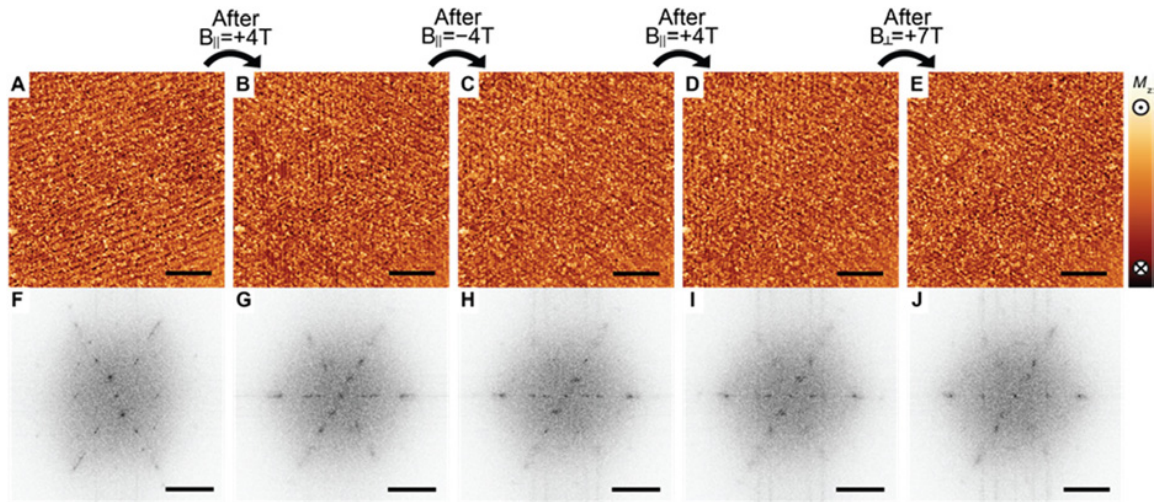

**Fig. S14. Aging of the spin-Q glass state in magnetic field with various directions.**
The measurements were taken at $T = 40$ mK, at a location close to the edge of a Nd island, which is presumably why the pristine state (**A**) has $Q$ states locked mostly along one high-symmetry direction. (**B-D**) Series of in-plane field aging experiments measured on the exact same area as the pristine state ($\tau_i = 10^3$ s for each magnetic field). (**E**) Out-of-plane aging of the same area ($\tau_i = 10^3$ s) for comparison with the aging experiments at $T = 1.3$ K (cf. Fig. 6) (scale bar = 30 nm, $I_t = 200$ pA). (**F-J**) Corresponding Q-space images of the images above (scale bar = 3 nm$^{-1}$).

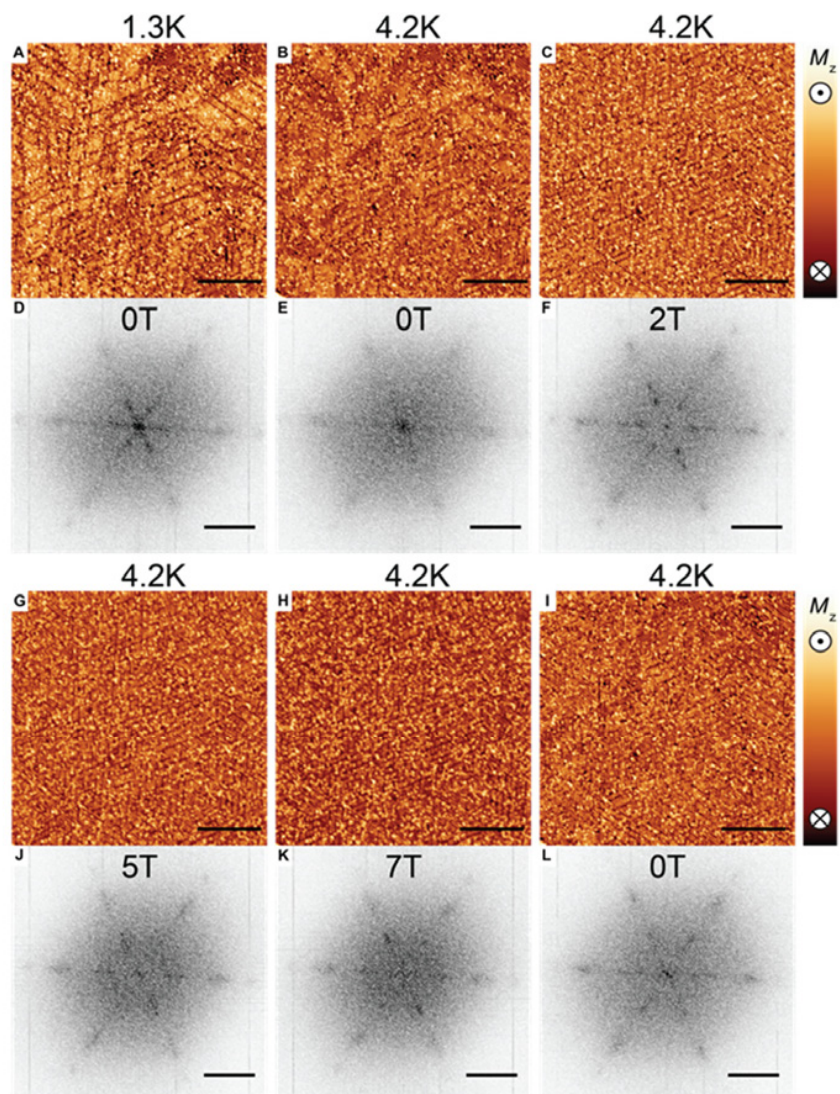

**Fig. S15. Temperature dependence of the Q-pockets in magnetic field.**
Magnetization images of the exact same area (**A**) at $T$ = 1.3 K in $B$ = 0 T and (**B**, **C**, **G-I**) at $T$ = 4.2 K in various out-of-plane magnetic fields (scale bar = 30 nm, $I_t$ = 200 pA). (**D-F**, **J-K**) Corresponding Q-space images of the images above (scale bar = 3 nm$^{-1}$). Warming up the surface from 1.3 K to 4.2 K results in depopulation of the $Q_A$ pocket. Similar out-of-plane magnetic field dependence and aging behavior is observed at $T$ = 4.2 K for the other Q-pockets. The magnetization image in (I), measured after magnetic field-dependent measurements, is the aged state of the pristine state in (B) ($t_i$ = 10$^5$ s at above 1 T).

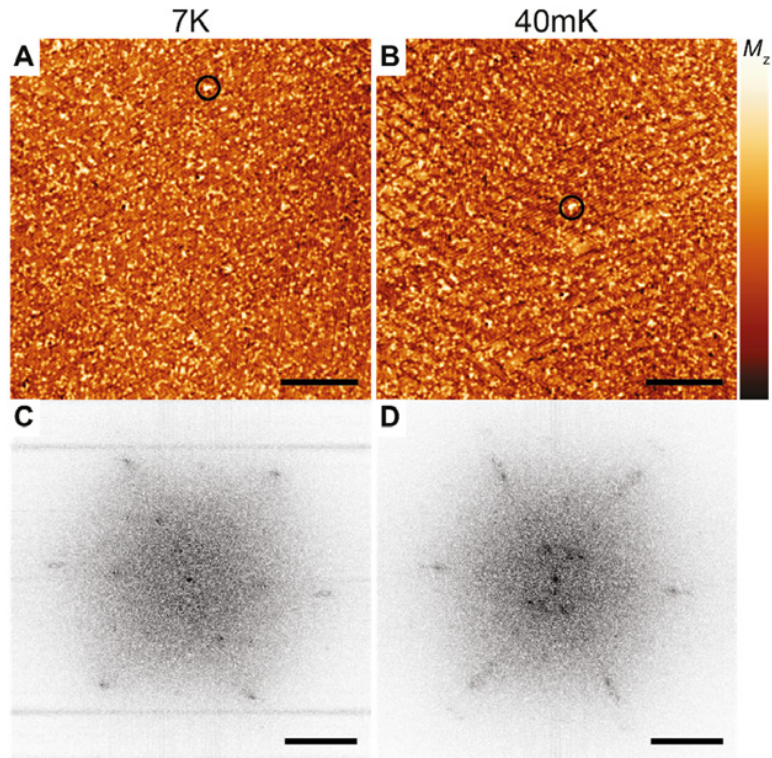

**Fig. S16. Temperature dependence of spin-Q glass state.**
Magnetization images of the same area (**A**) at $T$ = 7 K, (**B**) at $T$ = 40 mK (scale bar = 30 nm, $I_t$ = 200 pA), and (**C**, **D**) corresponding Q-space images of the images above (scale bar = 3 nm$^{-1}$). The surface shows fewer but more well-defined $Q$ states at $T$ = 7 K. The absence of spectral weight in the $Q_A$ pocket at higher temperature is similar to the observation at $T$ = 4.2 K.

**Table S1. Impurity concentration of the Nd source material.**

Elemental analysis (weight and atomic ppm, elements with <1 ppma not shown) as provided by Ames Laboratory. The sum of all rare-earth impurities (including Y) is 18.2 ppma, that of all 3$d$ elements 0.9 ppma.

| Element | Concentration (ppm wt.) | (ppma) | Element | Concentration (ppm wt.) | (ppma) |
|---|---|---|---|---|---|
| O  | 257   | 2316   | Pr | 3.5   | 3.6  |
| C  | 117   | 1704   | Y  | 2.2   | 3.6  |
| N  | 319   | 319    | Si | 0.6   | 3.1  |
| In | ≤43   | ≤54.0  | Dy | 3.3   | 2.9  |
| Al | 4     | 21.4   | Gd | 2.5   | 2.3  |
| Cd | <10   | <12.8  | Ce | 1.8   | 1.9  |
| Cs | <5    | <5.4   | Sm | 1.6   | 1.5  |
| F  | <0.5  | <3.8   | Ge | <0.5  | <1.0 |

**Table S2. Magnetic moments of Nd in different layers.**

Calculations show that the surface magnetic moment is restored to the bulk value after three layers. The 4f contribution to the magnetic moment is 2.454 µB, the rest is induced in the 6s and 5d shells.

| Layer | Magnetic moment (µ$_B$) | |
|---|---|---|
| | hexagonal surface termination | cubic surface termination |
| 1$^{st}$ | 2.89 | 2.91 |
| 2$^{nd}$ | 2.79 | 2.81 |
| Bulk cubic sites | 2.66 | |
| Bulk hexagonal sites | 2.67 | |